\documentclass{article}

\usepackage{arxiv}

\usepackage[utf8]{inputenc} 
\usepackage[T1]{fontenc}    
\usepackage{hyperref}       
\usepackage{url}            
\usepackage{booktabs}       
\usepackage{amsfonts}       
\usepackage{nicefrac}       
\usepackage{microtype}      
\usepackage{lipsum}		
\usepackage{graphicx}
\usepackage{natbib}
\usepackage{doi}
\usepackage{amsmath}
\usepackage{booktabs}
\usepackage{threeparttable}

\title{A Deep Learning-based time shift objective function for Full Waveform Inversion}

\author{ \href{https://orcid.org/0009-0005-5239-3955}{\includegraphics[scale=0.06]{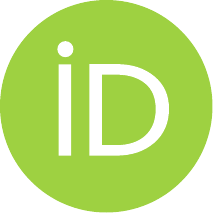}\hspace{1mm}Mustafa Alfarhan} \\
	KAUST\\
	Thuwal, Kingdom of Saudi Arabia\\
	\texttt{mustafa.alfarhan@kaust.edu.sa} \\
	\And
	{Fuqiang Chen} \\
	KAUST\\
	Thuwal, Kingdom of Saudi Arabia\\
	\texttt{fuqiang.chen@kaust.edu.sa} \\
        \And
	{George Turkiyyah} \\
	KAUST\\
	Thuwal, Kingdom of Saudi Arabia\\
	\texttt{george.turkiyyah@kaust.edu.sa} \\
        \And
	{David Keyes} \\
	KAUST\\
	Thuwal, Kingdom of Saudi Arabia\\
	\texttt{david.keyes@kaust.edu.sa} \\
        \And
	{Ivan Vasconcelos} \\
	Shearwater GeoServices\\
	Gatwick, United Kingdom\\
	\texttt{ivasconcelos@shearwatergeo.com} \\
        \And
	{Matteo Ravasi} \\
	Shearwater GeoServices\\
	Gatwick, United Kingdom\\
	\texttt{mravasi@shearwatergeo.com} \\
}

\renewcommand{\shorttitle}{Deep learning-based time shift FWI}

\hypersetup{
pdftitle={Deep learning-based time shift FWI},
pdfsubject={FWI},
pdfauthor={Mustafa Alfarhan},
pdfkeywords={Full waveform inversion, time-shift estimation, neural networks, automatic differentiation, adjoint sources},
}

\begin{document}
\maketitle

\begin{abstract}
Full Waveform Inversion (FWI) is a powerful technique for estimating high-resolution subsurface velocity models by minimizing the discrepancy between modeled and observed seismic data. However, the oscillatory nature of seismic waveforms makes point-wise discrepancy measures highly prone to cycle-skipping, especially when the initial velocity model is inadequate. To address this challenge, various alternative misfit functions have been proposed in the literature, each with unique strengths and limitations. Dynamic Time Warping (DTW) is a popular technique in signal processing for aligning time series using dynamic programming. While a differentiable variant of DTW has been recently proposed, its use in FWI is hindered by high-frequency artifacts in the adjoint source and the substantial computational cost of gradient evaluations. In this study, we propose a neural network-based approach to learn the time shifts that align two time series in a supervised manner. The trained network is then utilized to compare traces from observed and modeled seismic data, offering a stable and computationally efficient alternative to DTW. Furthermore, the inherent differentiability of neural networks via backpropagation enables seamless integration into the FWI framework as a misfit function. We validate this approach on two synthetic datasets, namely the Marmousi model and the Chevron blind test dataset, demonstrating in both cases a similar convergence behavior to that of SoftDWT whilst drastically reducing the computational time of the adjoint source calculation.
\end{abstract}


\section{Introduction}
Full Waveform Inversion (FWI) is a technique aimed at inverting seismic data for a high-resolution subsurface model by matching simulated and observed seismograms. Due to the oscillatory nature of seismic recordings and the use of gradient-based numerical optimization schemes, the highly nonlinear and ill-posed nature of FWI can lead to solutions falling into local minima, thus likely failing to recover a geologically plausible subsurface model. This problem, also known as cycle-skipping, arises when the simulated and observed seismograms are separated from each other by more than half a cycle. To alleviate the problem of cycle-skipping, hierarchical or multi-scale FWI is usually carried out~\citep{bunks1995multiscale}. In multi-scale FWI, inversion is initially performed using a low-frequency filtered version of the data. As the inverted model improves, the higher-frequency components are gradually introduced into the process. However, in practice, it is not always possible to access sufficiently low-frequency information to bootstrap the FWI process in the presence of a poor quality initial background model.

To mitigate cycle-skipping, an active area of research in FWI is the design of misfit functions that compare the kinematic component of the modeled and observed seismograms rather than their amplitudes, as is the case for the commonly used $L_2$ norm. Significant effort has been devoted to identifying misfit functions based on cross-correlation \citep{van2008velocity, van2010correlation,luo2011deconvolution, warner2016adaptive, zhu2016building, debens2017full, huang2017full}, minimizing the difference between the observed and calculated traveltimes using cross-correlation~\citep{10.1111/j.1365-246X.1991.tb06713.x, luo1991wave}, data envelope \citep{Bozdag2011845, CHI201436}, and the optimal transport distance~\citep{engquist2016optimal, metivier2016measuring, chen2018constructing, chen2018misfit, yang2018application, ramos2019long, kalita2019flux}. However, in practice, including any of these functionals in FWI comes with its own challenges. For example, some of them are sensitive to noise, unstable, require proper data normalization, and/or are computationally demanding.

A technique commonly used in signal processing to  compare two time series of possibly different lengths robustly is dynamic time warping (DTW) \citep{sakoe1978dynamic}. Despite its non-differentiability, the time shift estimate from the DTW algorithm has been successfully employed as an objective function in FWI~\citep{ma2013wave} and tomography~\citep{yang2014using} to mitigate the problem of encountering local minima. However, the discontinuous nature of this objective function can lead to sharp transitions in the adjoint source and the associated gradient~\citep{chen2021misfit}. A differentiable DTW method has been proposed by~\cite{cuturi2017soft} known as softDTW, which works by relaxing the minimum operator to a soft form. Recently,~\cite{chen2022cycle} applied differentiable softDTW with a penalization term to FWI to emphasize the traveltime differences. However, a misfit function derived from the time shift produced by the softDTW method suffers from two drawbacks: it can yield negative values, and its minimum is not necessarily obtained when the two time series are identical. To address these issues,~\cite{kalita2023soft} recently proposed using the divergence form of softDTW within FWI, motivated by the work of~\cite{blondel2021differentiable}. Although DTW and its variants have better convexity than other traveltime-based objective functions, they require significant hyperparameter tuning to reach reliable solutions. Furthermore, although dynamic programming can be used to compute the warping path efficiently, embedding this approach within FWI remains computationally expensive because a DTW problem must be solved for each trace at each FWI iteration.

Deep learning has recently emerged as an alternative solution for estimating the time shift or the time/space warping between two seismic datasets. \cite{doi:10.1190/segam2020-3427292.1} developed a deep learning workflow based on FlowNet to estimate optical flow between consecutive seismic slices, generating sequence stratigraphy without manual labeling and ensuring geological consistency through vertical-order constraints and cross-calibration of optical flow. Similarly, \cite{10.1190/geo2019-0724.1} introduced SeisFlowNet, a CNN-based approach to seismic image registration, which significantly improves the accuracy and robustness of traditional methods, particularly under noise, phase distortions, and frequency variations in the two datasets. More recently, \cite{9444561} adapted a modified VoxelMorph architecture based on UNet from medical imaging to 4-D seismic data, achieving detailed 3-D time shift estimations, enhancing the robustness and applicability of time-lapse seismic data analysis through a self-supervised learning approach. The application of such a UNet based architecture to seismic time shifts demonstrates the relevance of such multiscale detail-preserving architectures to structured shift-estimation tasks. Collectively, these studies demonstrate the significant potential of deep learning techniques in improving the mapping between seismic data. Building on these advances, we propose using deep learning to learn the time shift between traces of the modeled and observed seismic data, and therefore quantify their discrepancy as a metric for FWI. Since neural networks are differentiable, this approach offers a cost-effective alternative compared to DTW while still being potentially robust to cycle-skipping. Integrating the neural network into the FWI misfit function inherently produces smooth, well-behaved gradients, making it an attractive candidate for computing FWI adjoint sources. Furthermore, smoother, well-behaved gradients may facilitate faster and more stable convergence, although a formal analysis is beyond the scope of this study.

This paper is organized as follows. The Theory section presents the theoretical framework underlying our approach, including a detailed description of the neural network architecture and the training process used to estimate the time shift between two input signals. The Numerical Examples section provides a series of simulations to validate our method, outlining the setup and results of seismic survey experiments. The Discussion section offers an in-depth analysis of the results, highlighting the advantages of our approach, and identifying potential areas for improvement. Finally, the Conclusions section summarizes the key findings of this study.
\section{Theoretical Framework}
In this study, we integrate a neural network into the FWI workflow to efficiently and robustly compute the gradient associated with a time-shift-based objective function. The network is designed to predict the optimal time shift between traces of the shot gathers computed from an initial model and those from the observed shot gathers. This procedure consists of two stages: a training stage for the neural network in which the neural network learns the mapping, and a deployment stage where the network is applied and its gradient serves as the adjoint source of FWI. Moreover, since the primary focus of traveltime-based waveform inversion is on first arrivals or refractions, a preprocessing step is applied to the seismic data to mask any reflection prior to training. This step is necessary to ensure successful application of the network to data modeled in a smooth background velocity model, as reflections are typically absent or very weak.

\subsection{The training stage}
Incorporating a neural network into the FWI workflow to predict the time shift is most effective when using a pre-trained model. Since seismic data are oscillatory and contaminated with noise, a network trained in advance to predict the time shift between two time series offers clear advantages. We propose training the neural network in a supervised fashion; hence, the time shift must be known beforehand for each pair of traces in the training data. To generate the training data, we simulate seismic wavefields in a smooth background velocity model using a finite-difference modeling scheme. The shifted version of each synthetic trace is then obtained by time-domain interpolation, where the samples are displaced according to the generated time-shift function. In this work, we generate time shifts in a parametric manner and apply them to the reference data to produce the shifted data. Since the reference and shifted data differ by a known synthesized time shift, we can train a neural network in a supervised manner to estimate that shift.  A key component of this training process is the generation of realistic time shift functions that mimic those encountered in real-world scenarios.

\subsubsection{Time-shift generation}
We generate a smooth time shift function $\boldsymbol{\tau} (t)$ as the sum of multiple Gaussian functions:
\begin{equation}\label{eq1.1}
\boldsymbol{\tau} (t) = \sum_{i=0}^{N} \alpha_i e^{- \frac{(t - \mu_i)^2}{2 \sigma_i^2}}
\end{equation}
where $t$ denotes time, $N$ is the number of Gaussians, $\alpha_i$ is a scaling factor, and $\mu_i$ and $\sigma^2_i$ are the mean and variance of the $i^{th}$ Gaussian, respectively. Equation~\ref{eq1.1} enables the direct generation of per-sample time shifts for each seismic trace, without requiring any \textit{a priori} knowledge of the data. These time shifts are then applied to a reference seismic trace to produce an equivalent shifted trace, both of which are provided as input to our time-shift estimator neural network. 

\begin{figure}[!htb]
  \centering
  \includegraphics[width=0.5\textwidth]{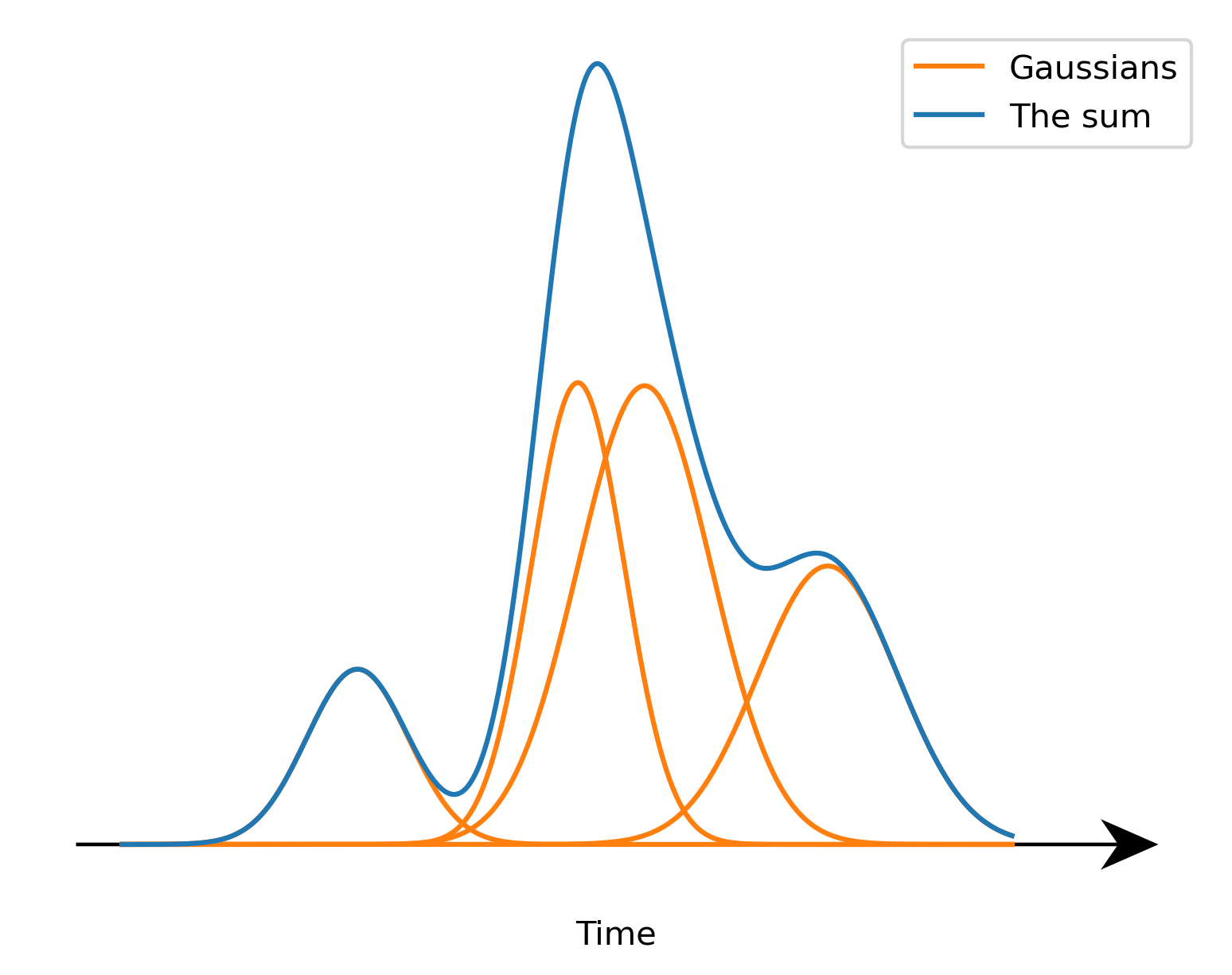}
  \caption{Example of the process used to generate time-shift functions.}
  \label{fig:timeshift}
\end{figure}

Our method offers several advantages for generating time shifts. First, the time-shift function is smooth, which facilitates the integration of the neural network within the FWI workflow. If the predicted time shift contains abrupt jumps, they appear in the adjoint source and contaminate the FWI gradient. Second, generating time shifts as the sum of Gaussians enables the creation of diverse training examples for training while ensuring smoothness. This smoothness preserves the temporal order of the signal so that time samples are not swapped. In this work, the number of Gaussians used to generate each time shift is selected between 3 and 5. The center of the Gaussian, $\mu$, is chosen to be between $20\%$ and $80\%$ of the length of the seismic trace, while $\sigma$ is chosen so that the resulting Gaussian does not produce sharp transitions. Figure~\ref{fig:timeshift} illustrates an example of a time shift generated with the proposed process. Finally, a post-processing step is applied after the time shift is introduced to avoid assigning shifts to regions of the trace where the seismic amplitude is zero, thereby preventing the network from encountering ill-posed training tasks. This is achieved by applying a mask to the ground-truth time shift with a binary window set to 1 over the shifted portion of the trace and 0 elsewhere, and using the masked time shift as the training label.

\subsubsection{Network and training details}\label{sec:network_design}
The network used to predict the time shift is a one-dimensional adaptation of the well-known UNet architecture~\citep{ronneberger2015u}. Our network consists of an encoder, a decoder, a bottleneck layer (bridging the encoder and decoder) and skip connections between corresponding encoder and decoder layers. The encoder is composed of multiple blocks, each containing a convolutional layer followed by a dropout layer and a ReLU activation function. This sequence is repeated, and the resulting feature maps are downsampled using a max-pooling layer. The process is applied four times, culminating in the bottleneck layer. The decoder mirrors the encoder's structure but replaces max-pooling with upsampling layers to restore the resolution of the input. To tailor the original UNet architecture for the time-shift prediction task, we introduce the following modifications:
\begin{itemize}
\item \textbf{Smoothing layer:} A smoothing layer is added after the final convolutional layer. Implemented as a convolutional layer with fixed, non-trainable parameters, this layer acts as implicit regularization. It prevents the network from generating sharp features or small fluctuations, while promoting smoother predictions.

\item \textbf{Increased kernel size:} We increase the kernel size in the convolutional layers from the conventional size of 3 to 11. A larger kernel enables the network to average over more time samples, which has a smoothing effect. In addition, it enables the network to capture long-range dependencies and extract more meaningful features.

\item \textbf{Removal of the first skip connection:} We remove the first skip connection from the network. The rationale is that early feature maps passed by the first skip connection to the decoder are close to the input and contain high-frequency content. Because time-shift prediction emphasizes low-frequency features to produce smooth outputs, removing this connection helps suppress high-frequency artifacts.
\end{itemize}
These modifications make the network more suitable for predicting smooth time shifts, thereby enhancing its performance for this specific application.

The input to the network consists of both the reference and shifted traces concatenated along the channel dimension, while the output is a single trace of the same size as the input traces, representing the time shift between them. Thus, the time shift predicted by the network is
\begin{equation}\label{eq1.1a}
\boldsymbol{\tau}_p (t) = \mathcal{NN}_{\theta} \big(d_1(t) \oplus d_2(t)\big)
\end{equation}
where $\mathcal{NN}_{\theta} (\cdot, \cdot)$ is the neural network parameterized by $\theta$, $d_1$ and $d_2$ are the reference and shifted traces, respectively, and $\oplus$ denotes the concatenation of $d_1$ and $d_2$ along the channel dimension. In practice, this means that $d_1$ and $d_2$ are stacked to form a multi-channel input tensor, with each trace occupying a separate channel. We use the mean absolute error between the true and predicted time shifts as the primary loss function to train our neural network:
\begin{equation}\label{eq1.1b}
L_{\text{MAE}} = \frac{1}{T} \| \boldsymbol{\tau}_\text{obs} (t) - \boldsymbol{\tau}_p (t)\|_1
\end{equation}
where $\boldsymbol{\tau}_\text{obs} (t)$ is the observed (ground-truth) time shift and $T$ is the length of the time axis. In addition, we apply the predicted time shift, $\boldsymbol{\tau}_p (t)$, to the reference trace, $d_1(t)$, to obtain a predicted shifted trace, $d_{2p}(t)$, which is then compared to the true shifted trace, $d_2(t)$. This comparison provides an auxiliary signal to improve gradient updates during training. We use the mean squared error (MSE) for this comparison: 
\begin{equation}\label{eq1.1c}
L_{\text{MSE}} = \frac{1}{T} \| d_2(t) - d_{2p}(t)\|^2_2
\end{equation}
The total loss for training is therefore defined as
\begin{equation}\label{eq1.1d}
L = L_{\text{MAE}} + \mu L_{\text{MSE}}
\end{equation}
where $\mu$ is a weighting factor that balances the losses. Empirically, we found that using the MAE loss for the time shift and the MSE for the data loss yields the best results.

\begin{figure}[!htb]
  \centering
  \includegraphics[width=\textwidth]{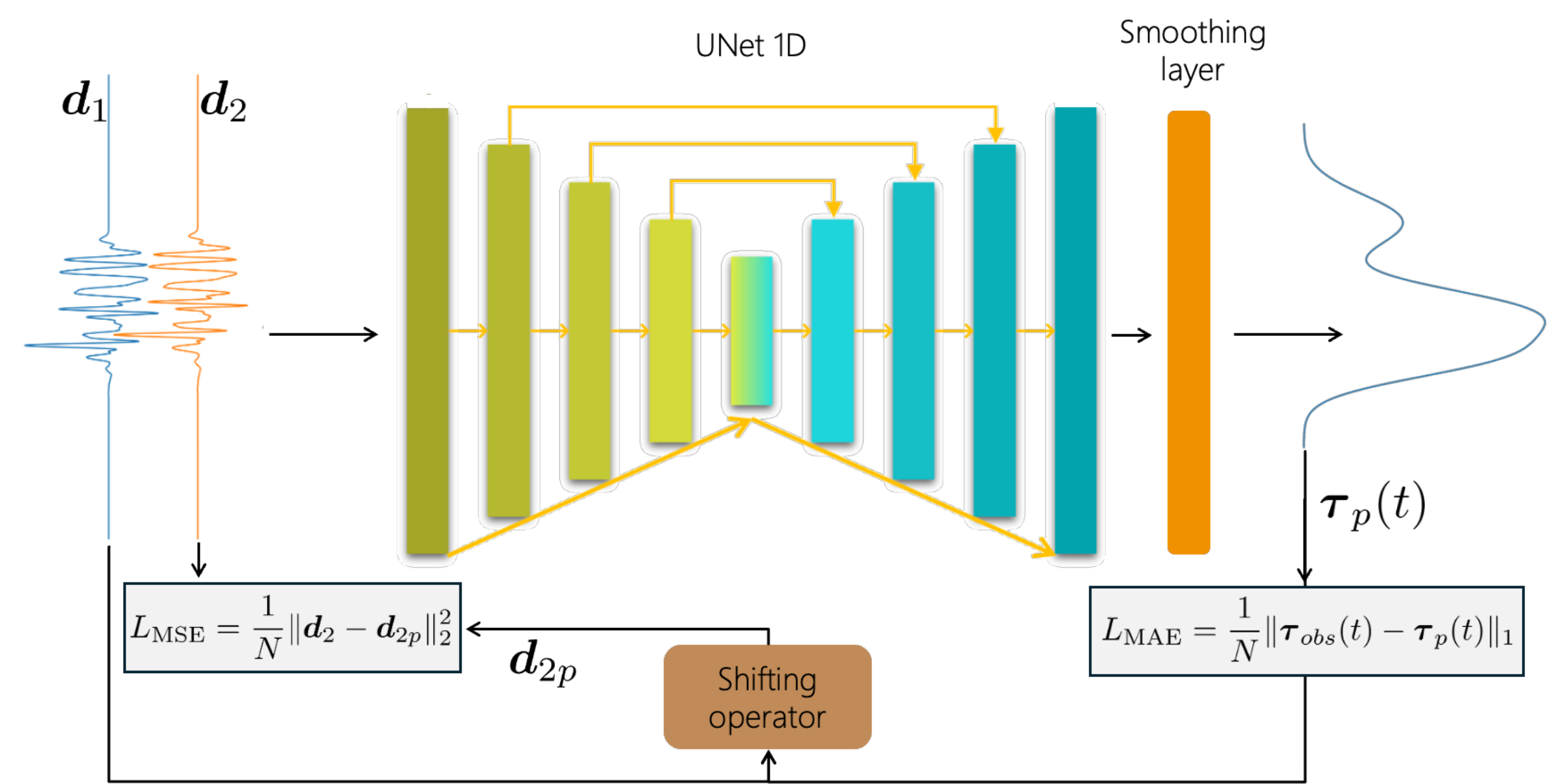}
  \caption{Workflow for training the neural network to predict time shifts.}
  \label{fig:train}
\end{figure}

In FWI, the reference trace and its shifted counterpart correspond to the modeled and observed data, respectively. After the initial training, the network was exposed to pairs of traces that differ only by time shift and amplitude scaling. However, in practice, the discrepancy between the modeled and observed data may involve additional factors beyond a simple time shift. To improve robustness, we fine-tune the network in a more realistic scenario, where the modeled data are denoted $d_1(t)$ and the observed data as $d_2(t)$, with potential differences in phase or waveform. This fine-tuning is performed using only the $L_{\text{MSE}}$ loss, which enables the network to estimate time shifts more reliably in the presence of such differences. Figure~\ref{fig:train} illustrates the training stage for time-shift prediction.

Although the network is one-dimensional, the batch dimension can be utilized as the receiver dimension of a shot gather during inference. This allows the time shift for the entire shot gather to be estimated in a single network pass, subject to the dimensions of shot gathers and the available GPU resources. Once the network has learned to predict the appropriate time shift between two datasets, it can be integrated into the FWI workflow. 

\subsection{The deployment stage}
Once the network has been trained, its output can be incorporated into the FWI misfit function: the predicted time shift represents the dissimilarity between the data modeled with a given initial model and the observed data. 
Using the $L_2$ norm as the misfit function for FWI, we define
\begin{equation}\label{eq1.3}
    \mathcal{J}(\mathbf{m}) = \frac{1}{2} \| \boldsymbol{\tau}_p(t)\|_2^2 = \sum_s \sum_r \frac{1}{2} \|  \mathcal{NN}_{\hat{\theta}} \big(d_{s,r}^\text{obs}(t) \oplus d_{s,r}(t)\big)\|_2^2
\end{equation}
where $\mathcal{NN}_{\hat{\theta}}$ denotes the trained neural network with parameters $\hat{\theta}$, $d_{s,r}^{\text{obs}}(t)$ is the observed trace at source $s$ and receiver $r$, and $d_{s,r}(t)$ is the corresponding modeled trace. Ideally, the misfit is zero when the time shift is zero; in other words, the modeled and observed data are perfectly aligned. 

\begin{figure}[!ht] 
\centering
\includegraphics[width=\textwidth]{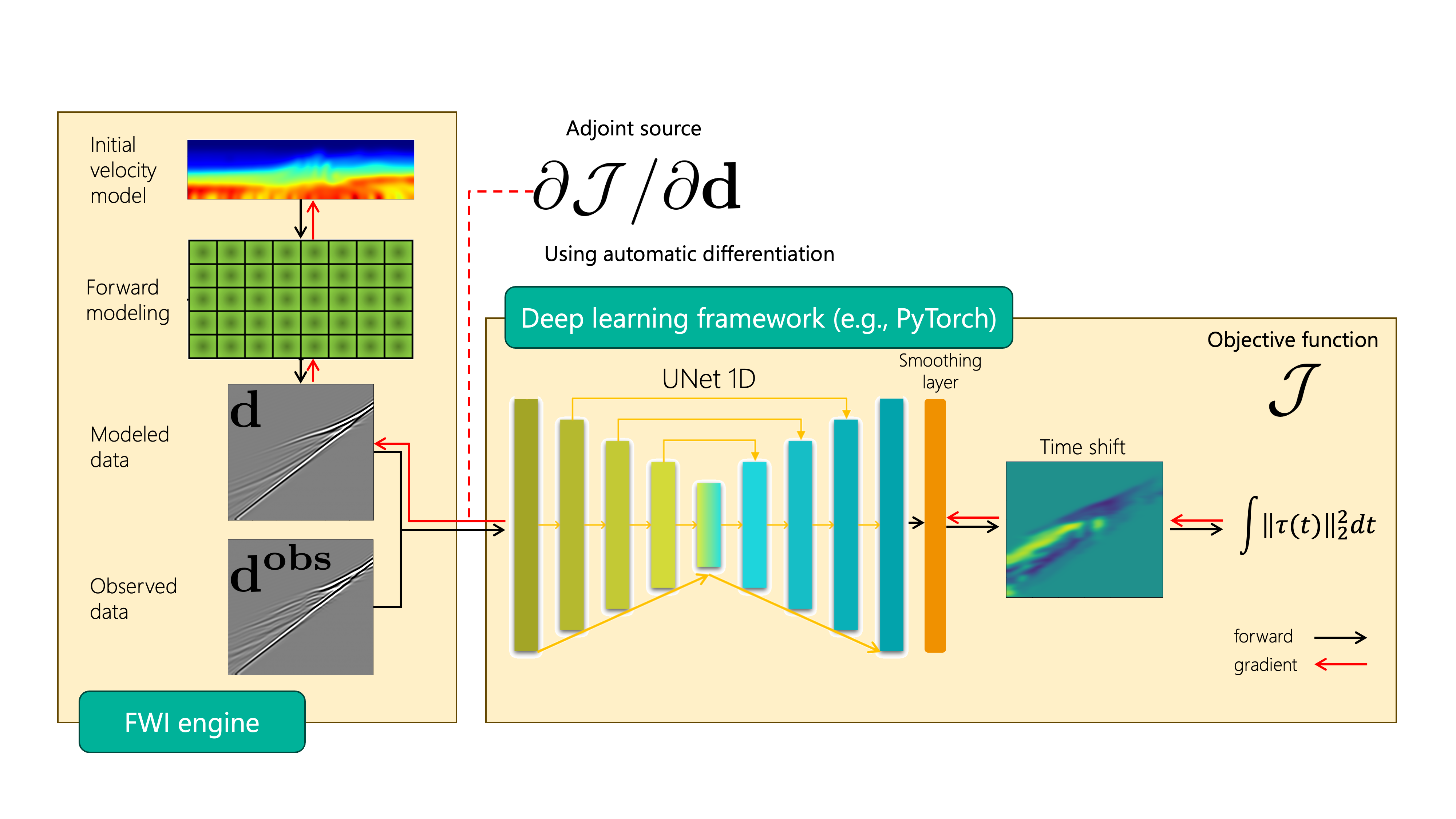} 
\caption{Schematic of the proposed method for approximating time shifts within the FWI workflow.}
\label{fig:fwi} 
\end{figure}

Let us now denote the modeled data $\mathbf{d}$ as
\begin{equation}\label{eq1.2}
    \mathbf{d} = \mathbf{P} \mathcal{L}(\mathbf{u}; \mathbf{m})
\end{equation}
where $\mathcal{L}$ is the forward modeling operator using the current model $\mathbf{m}$, $\mathbf{u}$ is the modeled wavefield, and $\mathbf{P}$ is a restriction operator that samples the source wavefield at the receiver locations. 
Using the adjoint state method~\citep{plessix2006review}, the derivative of the misfit function $\mathcal{J}$ with respect to the model parameters $\mathbf{m}$ can be express as
\begin{equation}\label{eq1.4a}
    \frac{\partial \mathcal{J}}{\partial \mathbf{m}} = \boldsymbol{\lambda}^T \frac{\partial \mathcal{L}}{\partial \mathbf{m}}
\end{equation}
where $\boldsymbol{\lambda}$ is the adjoint-state variable and $T$ denotes the transpose operator. For the acoustic wave equation, where $\mathbf{m} = \mathbf{c}$ (velocity), we obtain
\begin{equation}\label{eq1.4b}
    \frac{\partial \mathcal{L}}{\partial \mathbf{m}} = -\frac{2}{\mathbf{c}^3}\frac{\partial^2 \mathbf{u}}{\partial t^2}
\end{equation}
The adjoint state $\boldsymbol{\lambda}$ is solved from
\begin{equation}\label{eq1.4c}
    \frac{\partial \mathcal{J}}{\partial \mathbf{u}} = \boldsymbol{\lambda}^T \frac{\partial \mathcal{L}}{\partial \mathbf{u}} \Rightarrow \mathbf{P}^T \frac{\partial \mathcal{J}}{\partial \mathbf{d}} =  \mathcal{L}^* \boldsymbol{\lambda}
\end{equation}
where $\partial \mathcal{J}/\partial \mathbf{d}$ represents the adjoint source, obtained via backpropagation from the misfit function to the network input $\mathbf{d}$, and $\mathcal{L}^*$ is the adjoint wave equation operator (which, in the acoustic case, is identical to the forward operator). Finally, using a gradient-based optimization algorithm, the model is updated iteratively as
\begin{equation}\label{eq3.4}
    \mathbf{m}_{i+1} = \mathbf{m}_i - \beta \frac{\partial \mathcal{J}(\mathbf{m}_i)}{\partial \mathbf{m}}
\end{equation}
where the subscript $i$ indicates the current FWI iteration and $\beta$ is the step length. A schematic of the entire process is shown in Figure~\ref{fig:fwi}.

\section{Numerical Examples}
\subsection{Marmousi II}
The Marmousi II model~\citep{doi:10.1190/1.2172306} is a widely used synthetic benchmark for seismic imaging and FWI, known for its geological complexity. Compared to its predecessor, Marmousi II includes more complex geological structures, featuring highly variable sedimentary layers and a more prominent horizontal span. To evaluate our method, we designed a seismic survey simulating a seabed configuration. In this setup, single-component seismic receivers (hydrophones) are placed along the ocean floor, while seismic sources (air guns) are deployed just below the water's surface. The survey employs 60 sources uniformly distributed at a shallow depth of 10 meters. A total of 600 receivers are positioned at a depth of 450 meters (i.e., along the seafloor), starting 2000 meters from the left boundary and extending to 2000 meters short of the right boundary. This configuration is designed to mimic real-world seabed conditions as closely as possible.

\begin{figure}[!ht] 
\centering
\includegraphics[width=\textwidth]{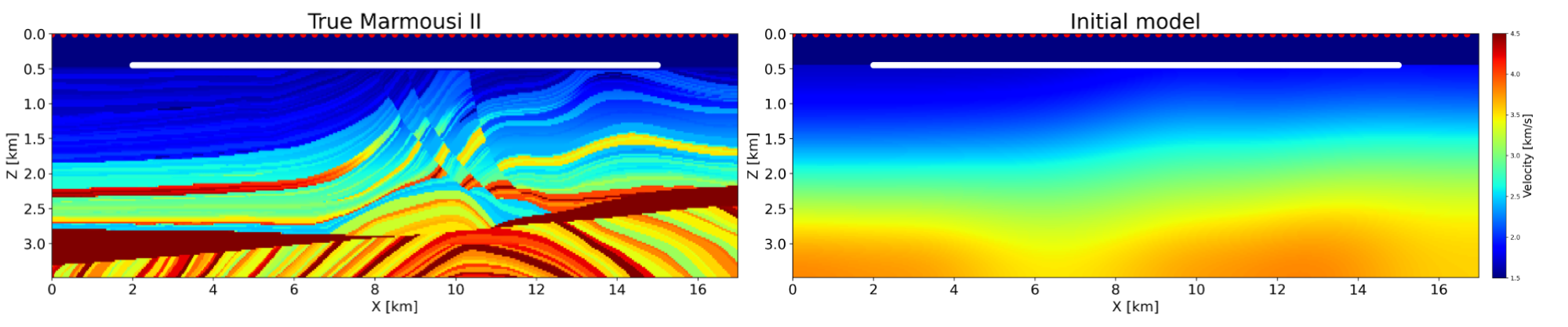} 
\caption{(a) The Marmousi II model with sources and receivers shown in red and white, respectively, and (b) the smoothed Marmousi II model.} 
\label{fig:marm} 
\end{figure}

The observed data are generated using a Ricker wavelet with a dominant frequency of 5 Hz and subsequently low-pass filtered with a cutoff frequency of 4 Hz. Figure~\ref{fig:marm} shows the Marmousi II model with the survey configuration and its smooth counterpart, which is used to obtain the modeled data and serves as an initial model for inversion. The velocity model used in our experiment is a spatially subsampled version of the original model by a factor of 20, using cubic spline interpolation for downsampling. Importantly, the physical dimensions, and hence the physics of wave propagation, are preserved. For brevity, we refer to the Marmousi II model simply as the Marmousi model throughout the remainder of this paper.

\begin{figure}[!ht] 
\centering
\includegraphics[width=\textwidth]{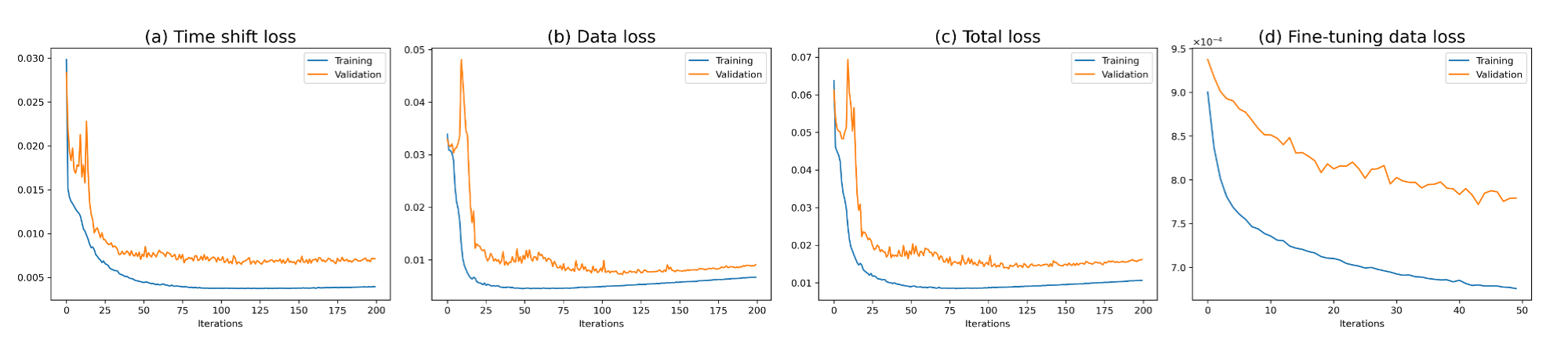} 
\caption{Training and validation loss curves for the neural network: (a) time-shift loss, (b) data loss, (c) total loss, and (d) fine-tuning loss.}
\label{fig:loss} 
\end{figure}

The shot gathers from the modeled data are split into 80\% training and 20\% validation sets to train the neural network. Seismic traces are randomly extracted from different shot gathers, paired with their shifted counterparts, and provided as inputs to the network. To simulate real-world scenarios in which observed data are contaminated with noise, Gaussian noise sampled from $\mathcal{N}(0, 1)$ is added to the shifted data. Using the known time shifts as labels, the network parameters are optimized with the AdamW optimizer, a learning rate of $10^{-4}$, and a cosine-annealing scheduler over 200 epochs. The training and validation datasets contain 14,400 and 3,600 traces, respectively, with a batch size of 256. Figure~\ref{fig:loss} shows the training and validation losses. Panels (a), (b), and (c) display the time shift loss, the data loss, and the total loss, respectively, all of which demonstrate that the network converges by the end of training. Panel (d) shows the fine-tuning loss, obtained from an additional 50 epochs of training with the Adam optimizer, using an initial learning rate of $10^{-6}$ and a cosine-annealing scheduler.

\begin{figure}[!ht] 
\centering
\includegraphics[width=\textwidth]{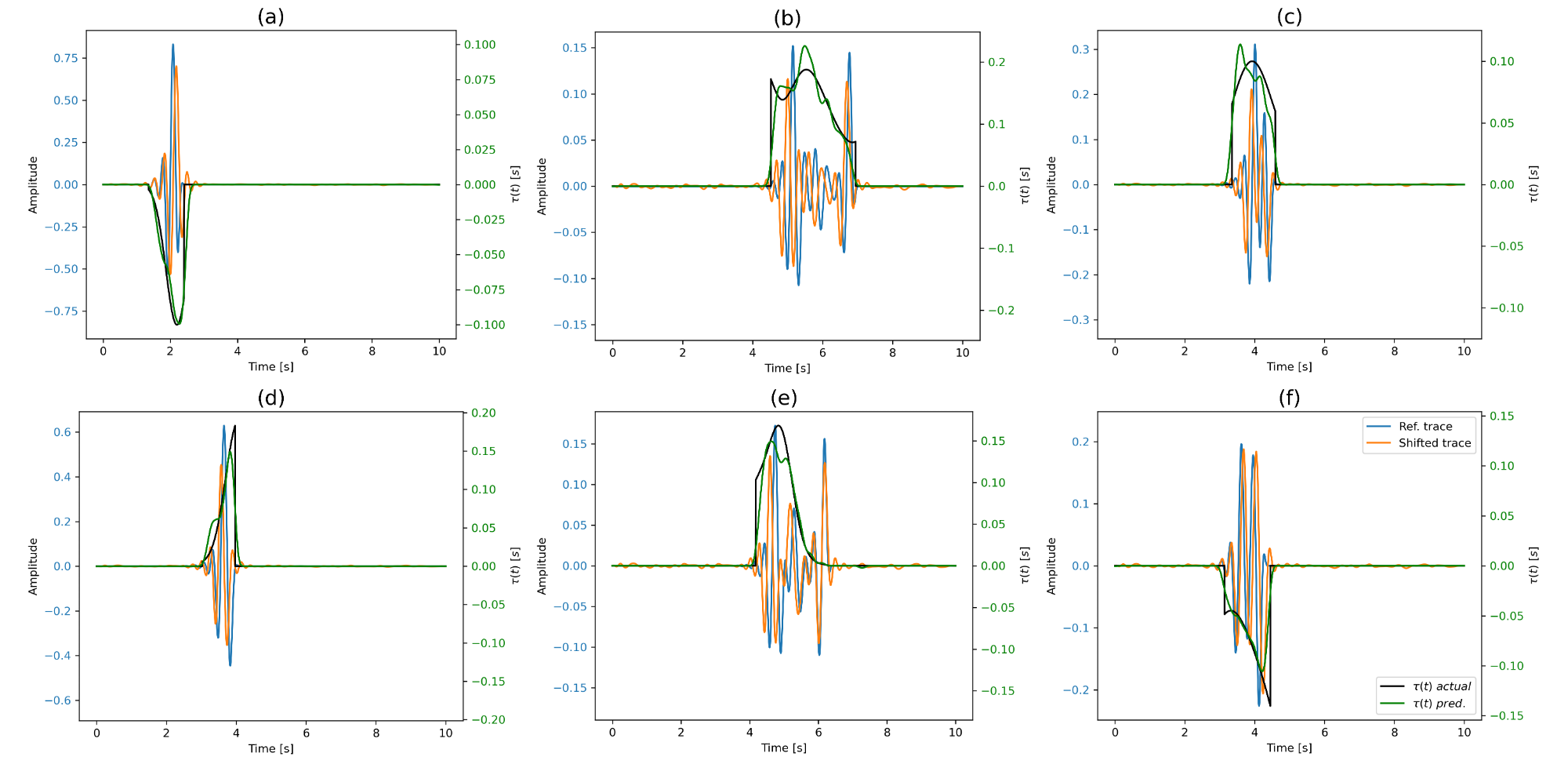} 
\caption{Comparison of predicted and ground-truth time shifts for six traces 
randomly selected from the validation dataset. The left y-axis represents the 
amplitude of the seismic traces, while the right y-axis represents the time 
shifts.}
\label{fig:pred} 
\end{figure}

\begin{figure}[!ht] 
\centering
\includegraphics[width=\textwidth]{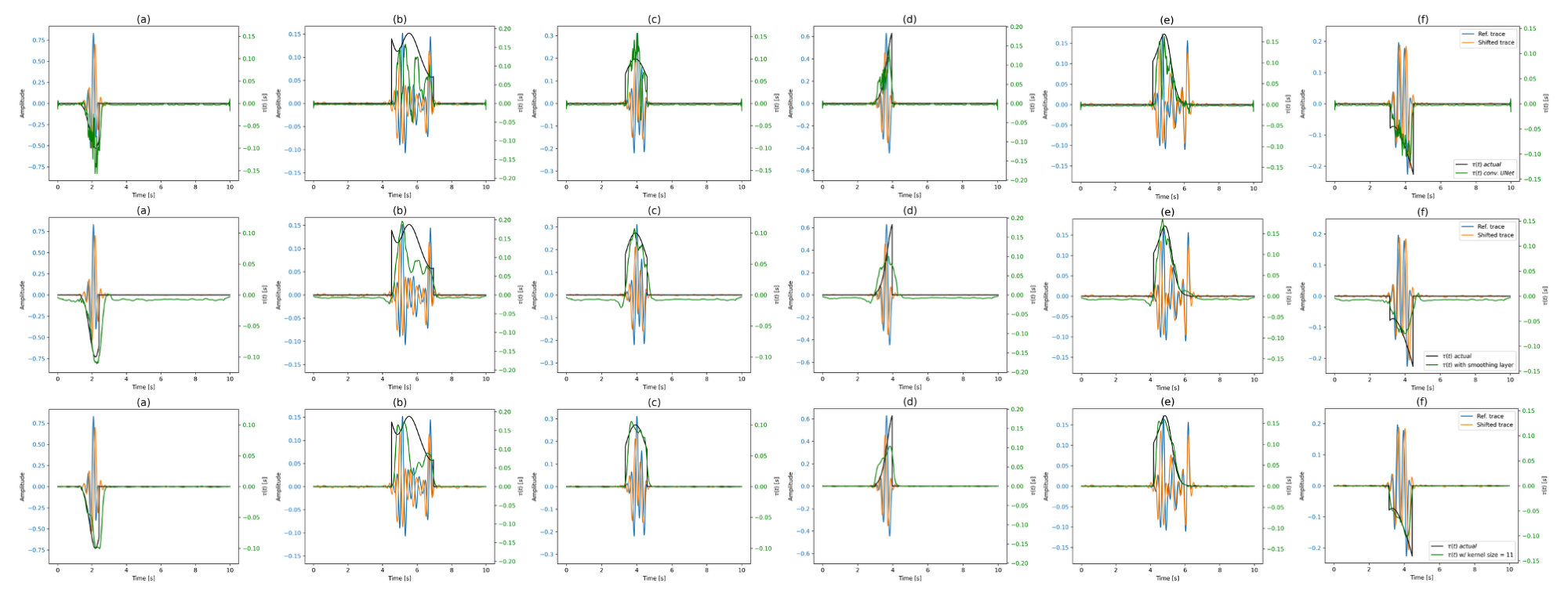} 
\caption{Ablation study of the network design choices for the time-shift prediction. First row: conventional UNet with kernel size 3. Second row: UNet with a smoothing layer added as the final layer. Third row: UNet with kernel size 11 plus smoothing layer.}
\label{fig:ablation} 
\end{figure}

For six traces randomly selected from the validation dataset, the predicted time shifts are shown in Figure~\ref{fig:pred} alongside the corresponding ground-truth shifts. The comparison demonstrates that the predicted time shifts closely match the actual values, indicating that the network successfully learned to capture the time shift between seismic traces. Figure~\ref{fig:ablation} illustrates the impact of the different network design choices presented in section~\ref{sec:network_design}. The first row shows predictions from a conventional UNet (kernel size = 3), where the estimated time shifts exhibit noticeable fluctuations. In the second row, when a smoothing layer is added at the end of the UNet architecture, the stability of the predictions is substantially improved. In the third row, when all convolutional layers employ larger kernel of size 11 alongside the smoothing layer, the remaining fluctuations are the model yields smoother time‐shift estimates. Finally, the benefit of removing the first skip connection has already been demonstrated in Figure~\ref{fig:pred}.

\begin{figure}[!ht] 
\centering
\includegraphics[width=0.8\textwidth]{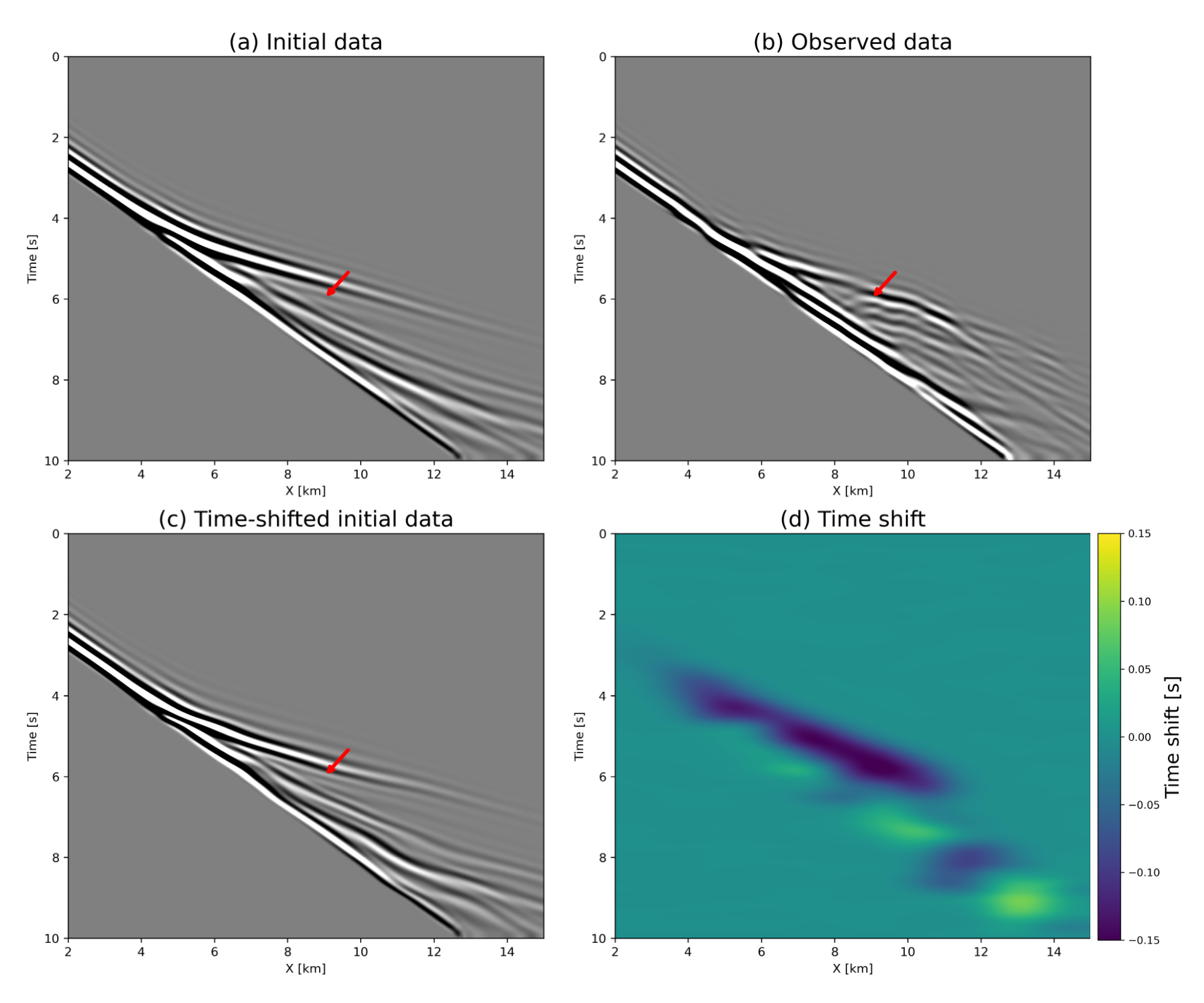} 
\caption{Application of the proposed method to a full shot gather. (a) Modeled data generated from the initial smooth velocity model, (b) the corresponding observed data, (c) the modeled data after applying the predicted time shifts, and (d) the estimated time-shift field.} 
\label{fig:pred_valid} 
\end{figure}

Finally, we test the network capabilities to predict actual times shifts in the context of FWI: instead of using observed (or modeled) data and a shifted version as inputs, the network takes as inputs the data modeled from the initial velocity model in Figure~\ref{fig:marm}(b) and observed data. Note that during training, the network has been exposed to only a portion of the data, whereas during testing, it must handle all available data. Moreover, the network was trained on data modeled from the initial velocity model, while during FWI it must process data at various stages of the inversion process -- i.e., which originate from intermediate velocity models. Thus, this experiment serves as a test of the network’s generalization capability. The task of the network is to predict the time shift that best aligns the modeled data with the observed data. Consistent with the training procedure, the modeled and observed data are provided to the network as inputs through the first and second channels in the input, respectively. The network has learned to determine the time shift that aligns the reference trace (modeled) with the shifted trace (observed).  Although the network accepts one-dimensional inputs, it can predict the time shift for an entire shot gather by treating the batch dimension as the receiver dimension. while this approach is theoretically equivalent to estimating time shifts for each trace individually, it provides a more efficient means of calculating multiple time shifts in a single pass of the network. Figures~\ref{fig:pred_valid}(a) and (b) show a modeled shot gather and the corresponding observed shot gather, respectively. The shifted version of the modeled shot gather and the estimated time shift are shown in Figures~\ref{fig:pred_valid}(c) and (d), respectively. An additional Gaussian smoothing filter is applied along the receiver dimension to further enhance the spatial coherency of the estimated time shifts. Finally, the arrows in panels (a)-(c) indicate how one of the key arrivals in the observed data is better aligned with the shifted gather than with the initial modeled gather.

After verifying the plausibility of the time shift predicted by the network between the modeled and observed data, we integrate the fine-tuned network into the FWI workflow (Figure~\ref {fig:fwi}) and perform FWI using the L-BFGS optimizer~\citep{liu1989limited}. In addition to comparing our approach to the conventional $L_2$ norm, we also compare it with the divergence form of softDTW, using a regularization parameter $\gamma = 10$ (hereafter referred to as SoftDTW for brevity). The inversion is conducted in three stages. First, we perform refraction FWI (excluding reflections) up to a maximum frequency of 4 Hz for 40 iterations. The reflections are suppressed by a mask based on the traveltime of the direct arrivals. Next, we carry out reflection FWI (the entire data with the $L_2$ norm) for 100 iterations. Finally, we run a second stage of reflection FWI with a maximum frequency of 7 Hz for 100 iterations. The iteration counts given above represent the maximum number of iterations; the actual number is determined by the stopping criteria of the L-BFGS optimizer, which may stop the process earlier.
\begin{figure}[!ht] 
\centering
\includegraphics[width=\textwidth]{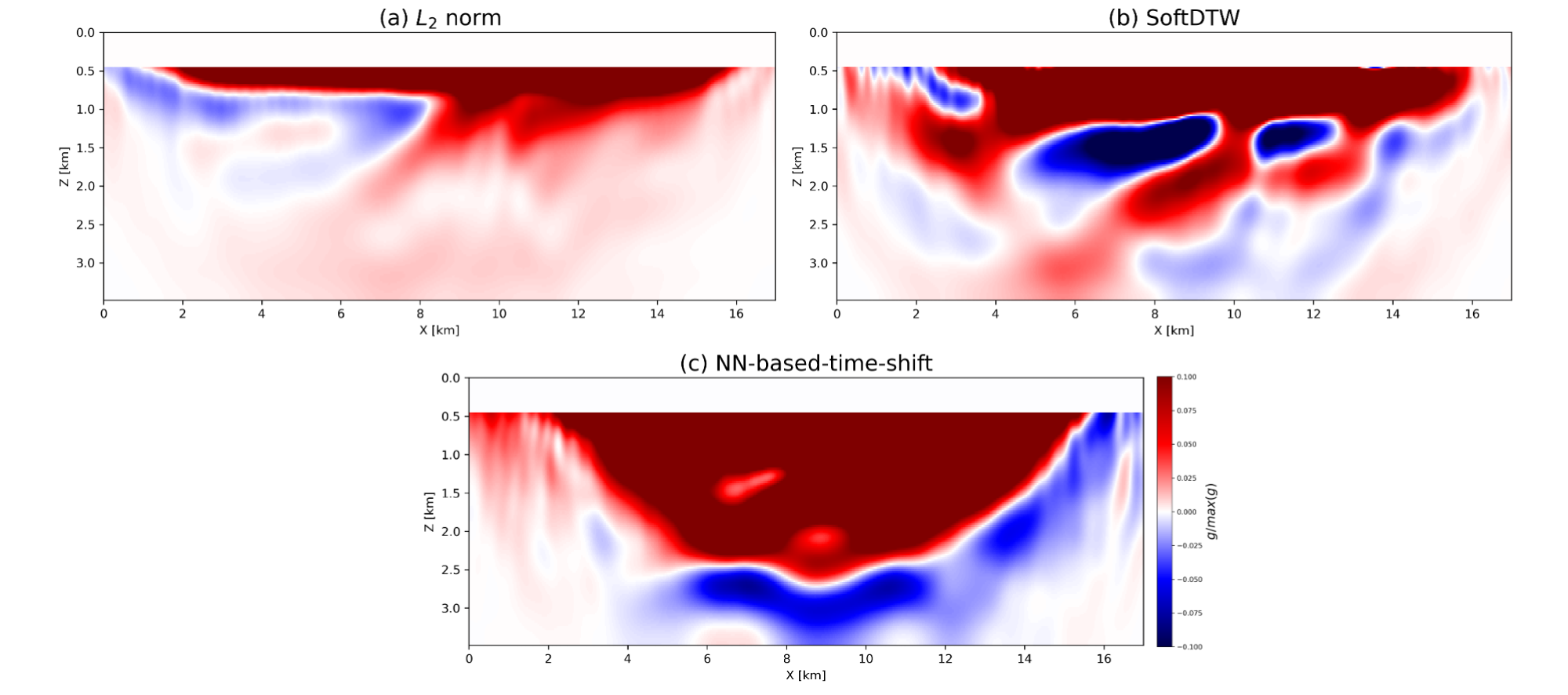} 
\caption{First-iteration FWI gradients obtained using different misfit functions: 
(a) conventional $L_2$ norm, (b) SoftDTW with $\gamma = 10$, and (c) the proposed 
time-shift–based approach.} 
\label{fig:grad} 
\end{figure}
\begin{figure}[!ht] 
\centering
\includegraphics[width=0.8\textwidth]{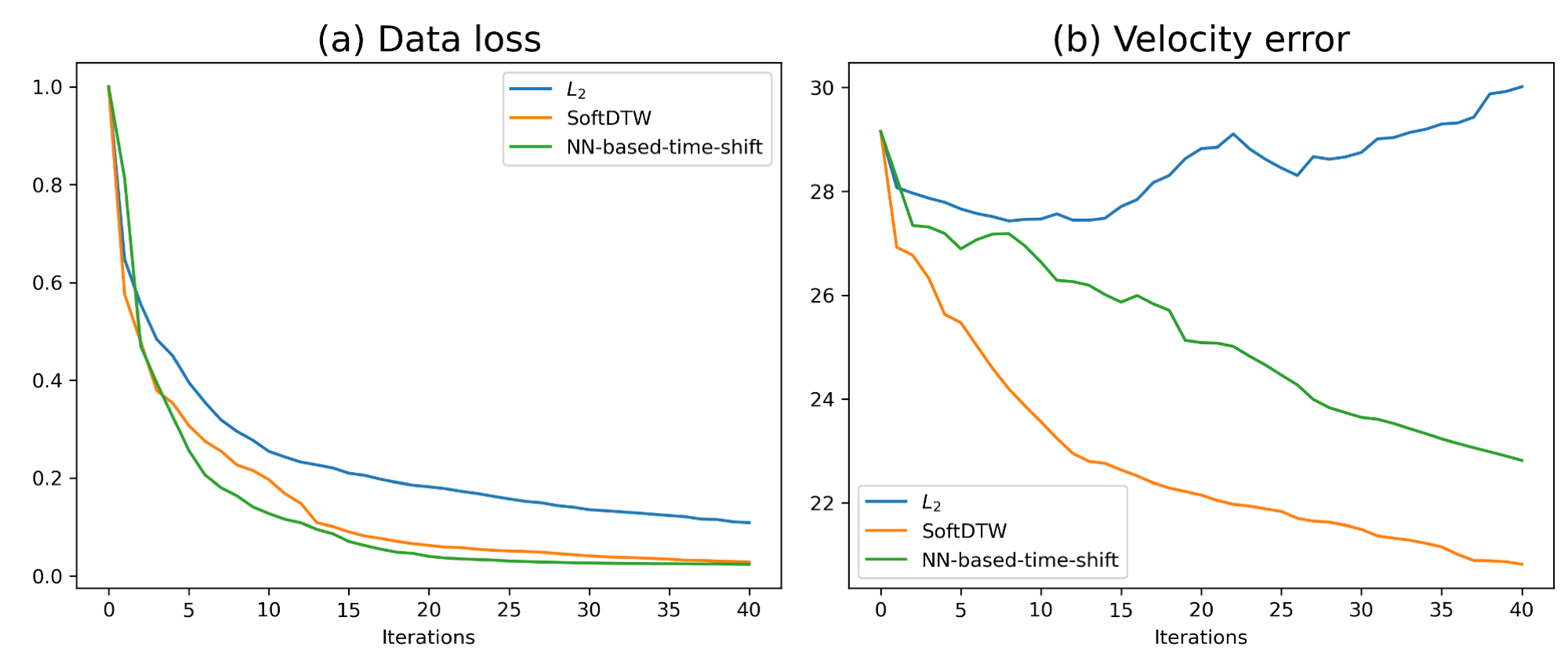} 
\caption{Performance of refraction FWI on the Marmousi model. 
(a) Data loss as a function of iteration, comparing different misfit functions, 
and (b) velocity error relative to the true model.} 
\label{fig:loss_err} 
\end{figure}
\begin{figure}[!ht] 
\centering
\includegraphics[width=\textwidth]{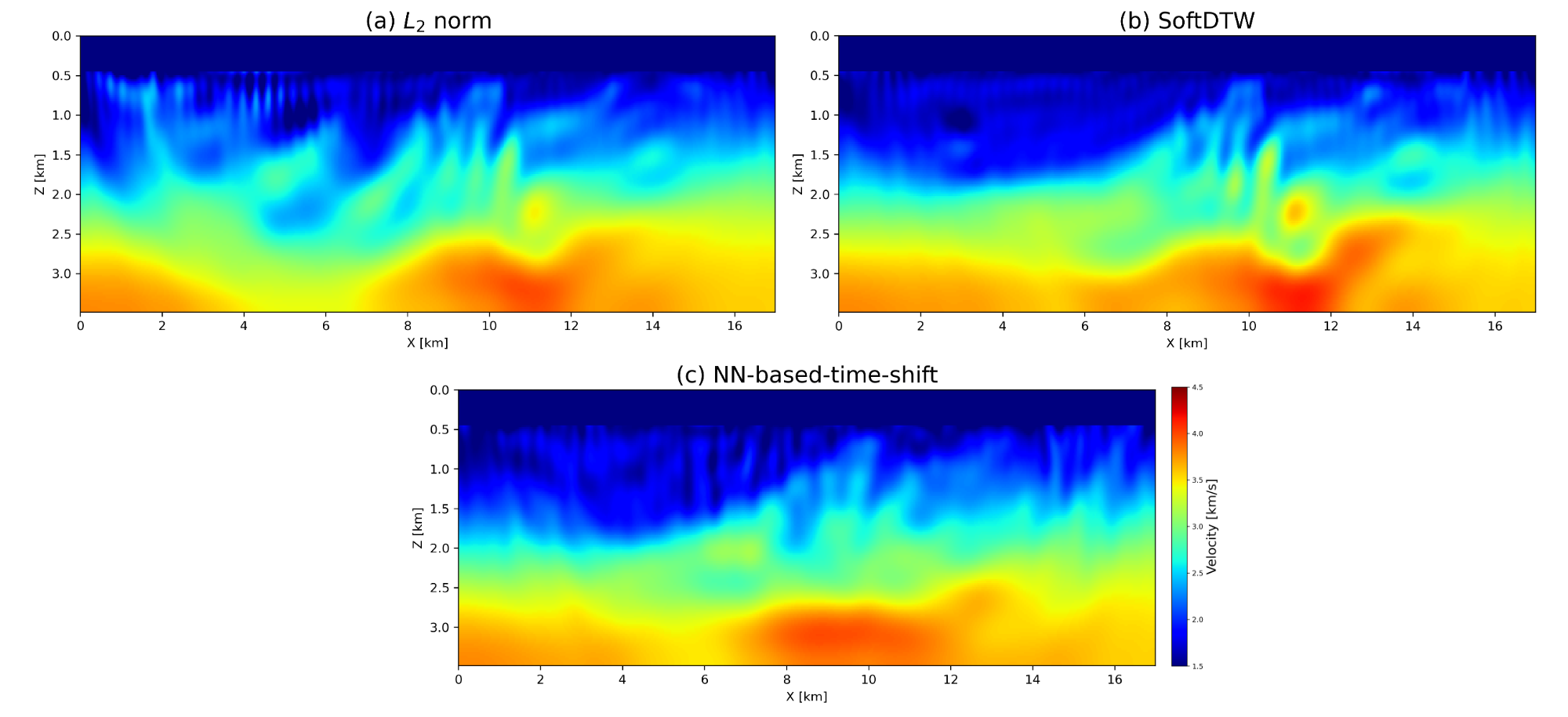} 
\caption{Final inverted velocity models for the Marmousi benchmark obtained with different misfit functions: (a) conventional $L_2$ norm, (b) SoftDTW with $\gamma = 10$, and (c) the proposed time-shift–based approach. } 
\label{fig:vel} 
\end{figure}
Figure~\ref{fig:grad}(a)-(c) show the first-iteration FWI gradient computed with the conventional $L_2$ norm, SoftDTW, and our proposed approach, respectively. All three gradients capture the major upper structures of the Marmousi model. Figures~\ref{fig:loss_err}(a) and \ref{fig:loss_err}(b) show the normalized data loss (relative to its maximum) and the velocity error, defined as $\|(v_k - v_{true})/v_{true}\|$, where $v_{true}$ is the true velocity and $v_k$ is the updated velocity at iteration $k$, for the three approaches. As expected, all data losses decrease with iteration; however, our approach achieves a faster reduction and reaches a lower loss than both the $L_2$ norm and SoftDTW. 

The velocity error, which provides a more reliable measure of convergence, tells a different story. The velocity update obtained with the $L_2$ norm diverges from the true solution. In contrast, both the SoftDTW and our time-shift-NN based approach converge toward the true solution. Importantly, our method exhibits simultaneous decreases in both data loss and velocity error, demonstrating that the predicted time shifts are meaningful and effectively align the modeled data with the observed data. 

\begin{figure}[!ht] 
\centering
\includegraphics[width=0.8\textwidth]{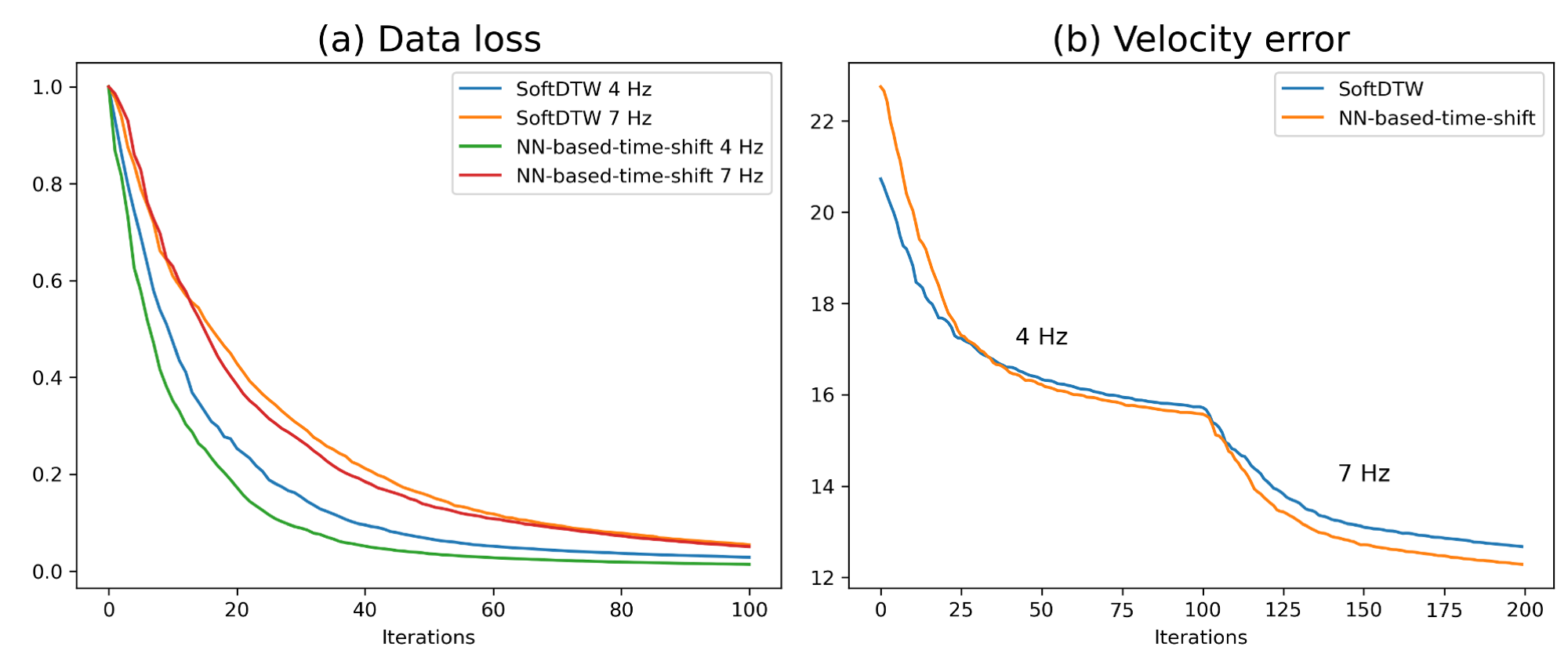} 
\caption{Performance of reflection FWI on the Marmousi model. (a) Data loss as a function of iteration, normalized by the maximum, and (b) velocity error relative to the true model.} 
\label{fig:loss_err_marm2} 
\end{figure}

\begin{figure}[!ht] 
\centering
\includegraphics[width=\textwidth]{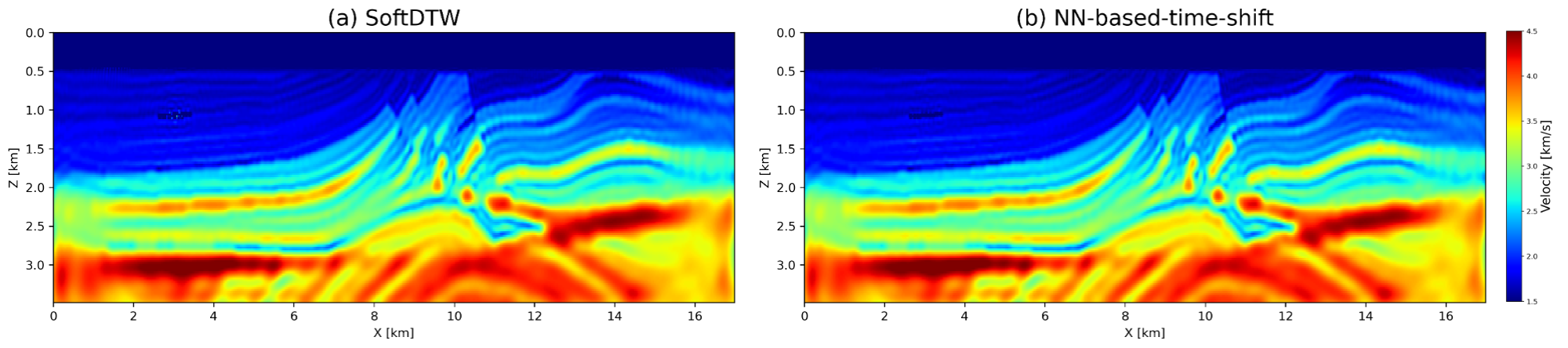} 
\caption{Final inverted velocity models for reflection FWI on the Marmousi benchmark: (a) SoftDTW with $\gamma = 10$ and (b) the proposed time-shift–based approach.} 
\label{fig:vel_marm2} 
\end{figure}

The final inverted velocity models for the three approaches are shown in Figure~\ref{fig:vel}. Our approach produces smoother updates, particularly in the shallower region, with fewer artifacts compared to both the $L_2$ norm and SoftDTW. The performance of FWI when using the SoftDTW- and NN-based models as starting points for the 4- and 7-Hz frequency bands is presented in Figure~\ref{fig:loss_err_marm2}. Although the starting model of our approach has a slightly higher velocity error, both the data loss and velocity error reach lower values than those obtained with SoftDTW. The final inverted velocity models for this experiment are shown in Figure~\ref{fig:vel_marm2}.
\subsection{SEG 2014 Chevron FWI Synthetic}
\begin{figure}[!ht] 
\centering
\includegraphics[width=\textwidth]{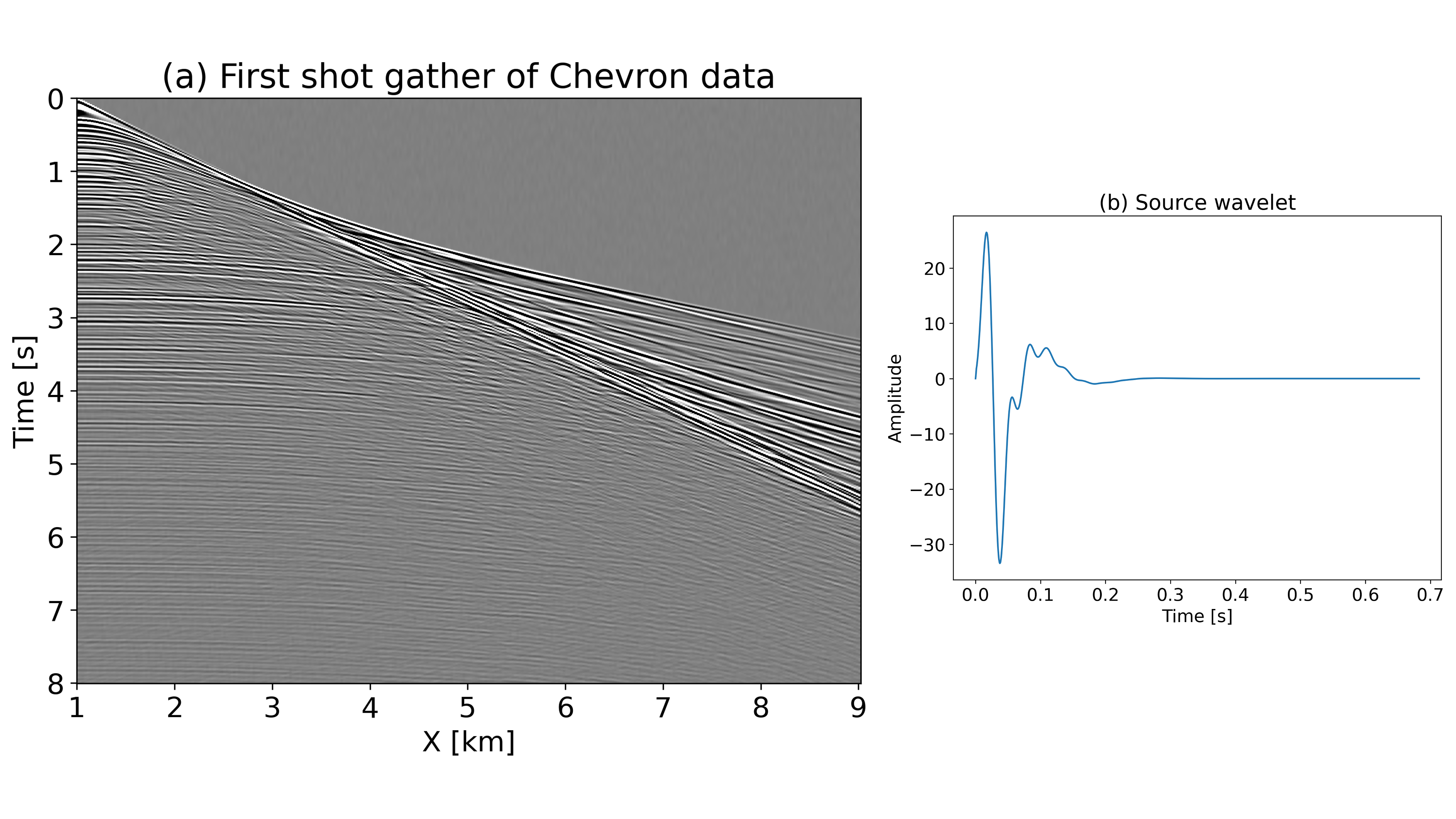} 
\caption{(a) First shot gather from the Chevron dataset and (b) the provided source wavelet.}
\label{fig:chevron_data} 
\end{figure}

\begin{figure}[!ht] 
\centering
\includegraphics[width=\textwidth]{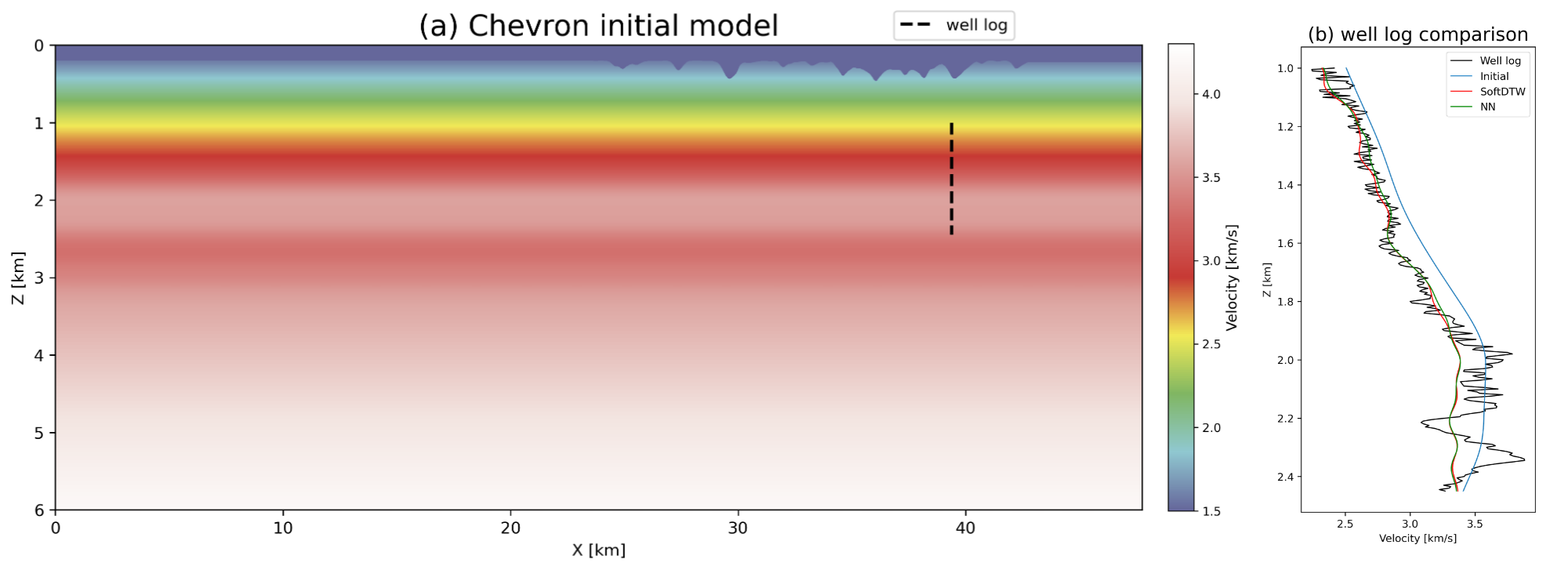} 
\caption{Chevron initial velocity model with the provided well log (trajectory at $x=39.375\ \text{km}$, $z=[1.0,\ 2.45]\ \text{km}$, shown in black). (a) Initial velocity model and (b) velocity profile along the well trajectory.}
\label{fig:chevron_vp_init} 
\end{figure}

Chevron released a synthetic marine dataset as a benchmark for blind tests during the SEG 2014 workshop. The dataset was generated using the 2D elastic isotropic wave equation from a model that features complex geological structures, including a free-surface boundary at the top. The recorded data are contaminated by both low- and high-frequency noise, making the inversion problem more realistic and challenging. The dataset consists of 1,600 shots acquired by 321 hydrophones, each recording 8 s of data sampled at 4 ms intervals. Additionally, Chevron provided a far-field wavelet, a single well log, and an initial velocity model to support the inversion process. Figure~\ref{fig:chevron_data} shows the first shot gather and the provided wavelet, while Figure~\ref{fig:chevron_vp_init}(a) shows the initial velocity model.

The acquisition geometry of the dataset consists of a streamer setup with a source-receiver offset starting at 1 km and both sources and receivers positioned at a depth of 15 m. The geometry features a 25 m spacing between elements and a maximum offset of 8 km. The model spans 6 km in depth and 47.75 km laterally. The Chevron dataset, along with its initial velocity model, poses significant challenges, as conventional $L_2$ norm-based FWI methods often struggle to achieve accurate results. To assess the effectiveness of our proposed approach, we performed FWI on a subset of 256 shots, starting from the first shot and taking every sixth shot up to the 1,531st shot. The training procedure for the Chevron dataset is similar to those used for the Marmousi model as described in the Theoretical Framework section.

We begin the inversion process using refraction data only, progressing through three stages with maximum frequencies of 4 Hz, 5 Hz, and 6 Hz. The choice of stepping through different frequency bands is motivated by the fact that the data is cycle skipped and therefore a more accurate tomographic model is required prior to switching to reflections (and an $L_2$ loss term). The workflow then continues with reflection data in three stages, with maximum frequencies of 7 Hz, 10 Hz, and 15 Hz. Similar to the Marmousi example, we perform FWI using SoftDTW and our Time-Shift-NN based misfit for refraction data and the $L_2$ norm for reflection data.

\begin{figure}[!ht] 
\centering
\includegraphics[width=\textwidth]{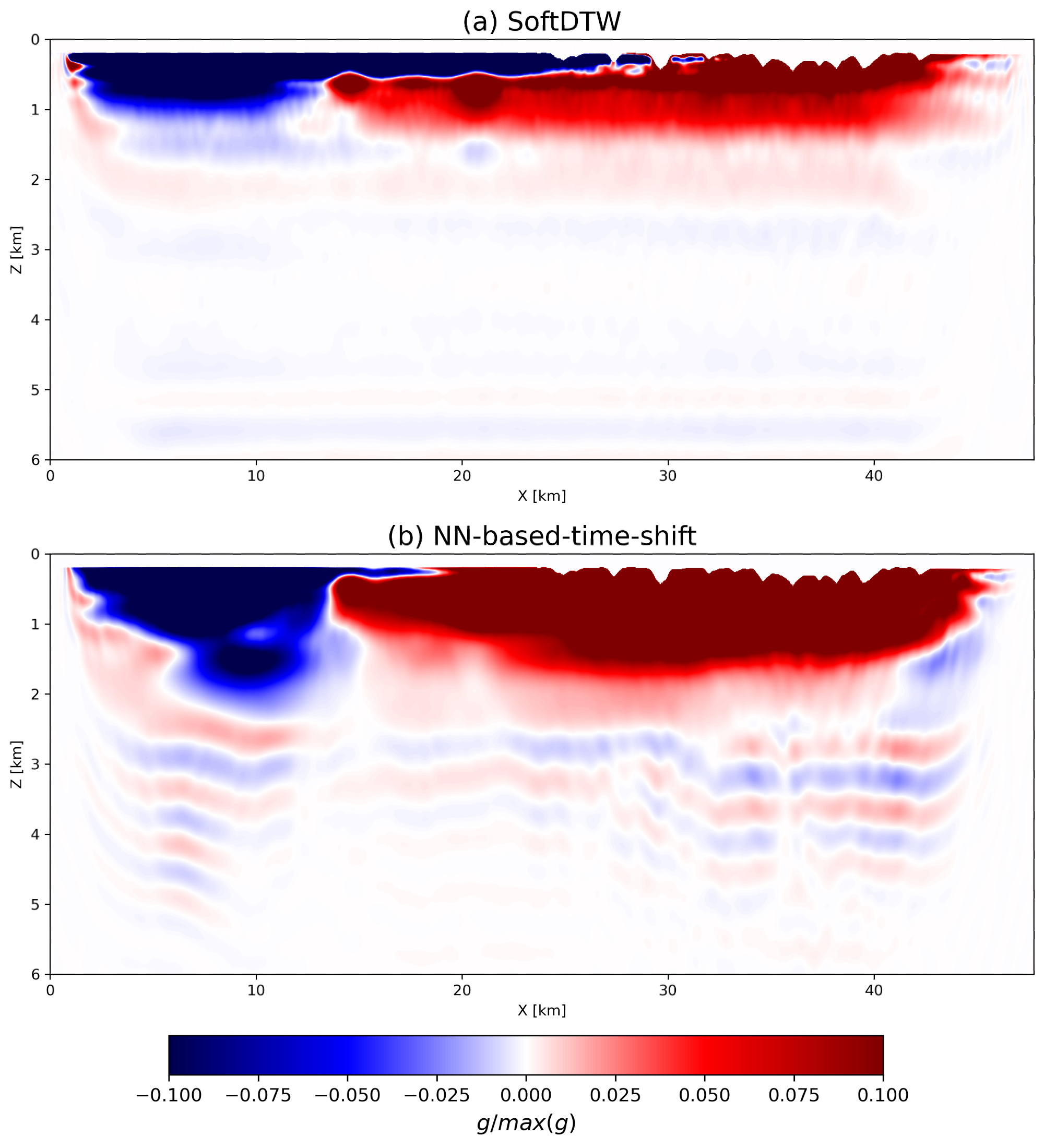} 
\caption{First-iteration FWI gradient at 4 Hz for the Chevron dataset using (a) SoftDTW and (b) our misfit.}
\label{fig:chevron_grad} 
\end{figure}

\begin{figure}[!ht] 
\centering
\includegraphics[width=\textwidth]{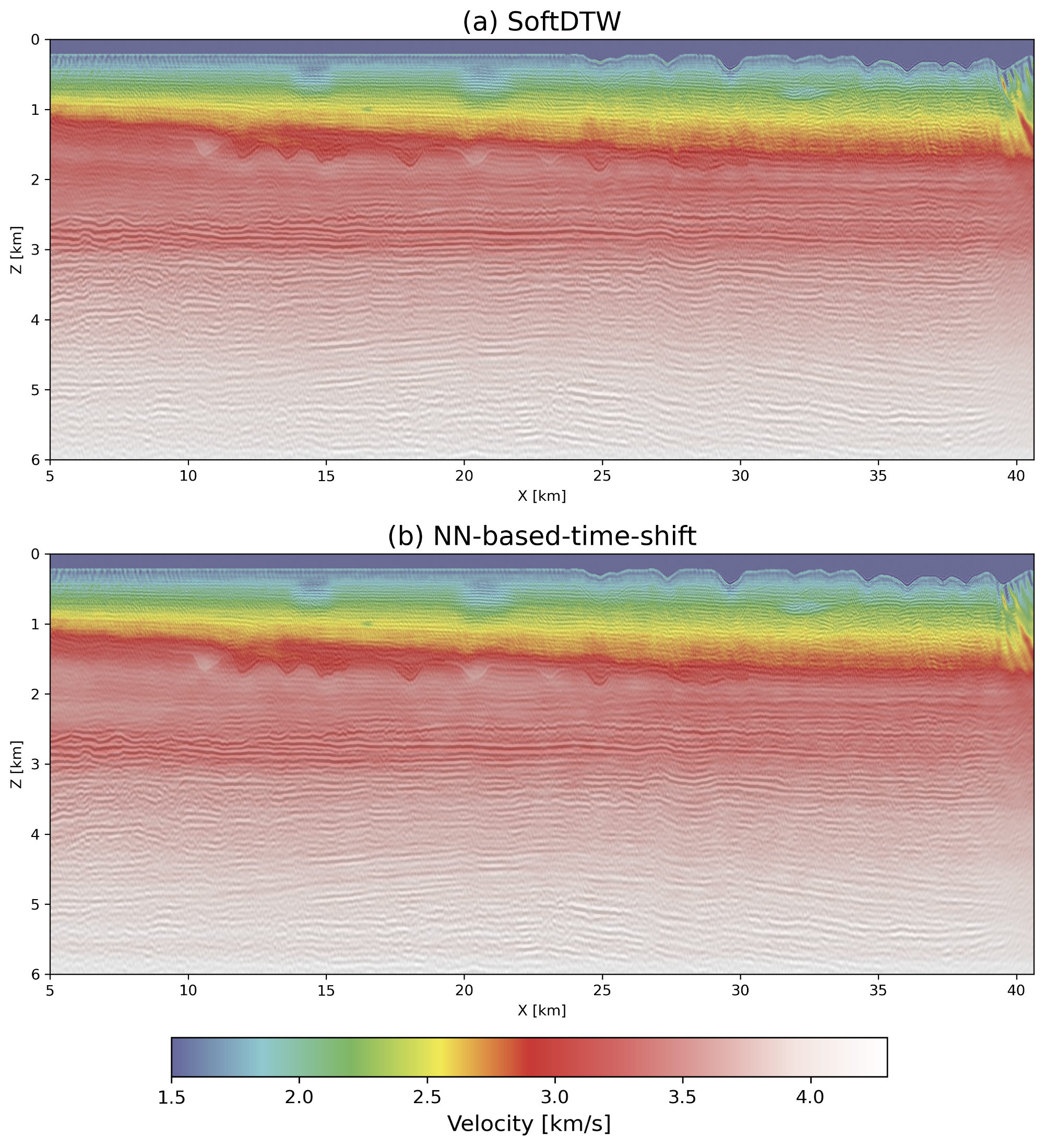} 
\caption{Final inverted Chevron velocity models using (a) SoftDTW and (b) our misfit.}
\label{fig:chevron_vp} 
\end{figure}

\begin{figure}[!ht] 
\centering
\includegraphics[width=\textwidth]{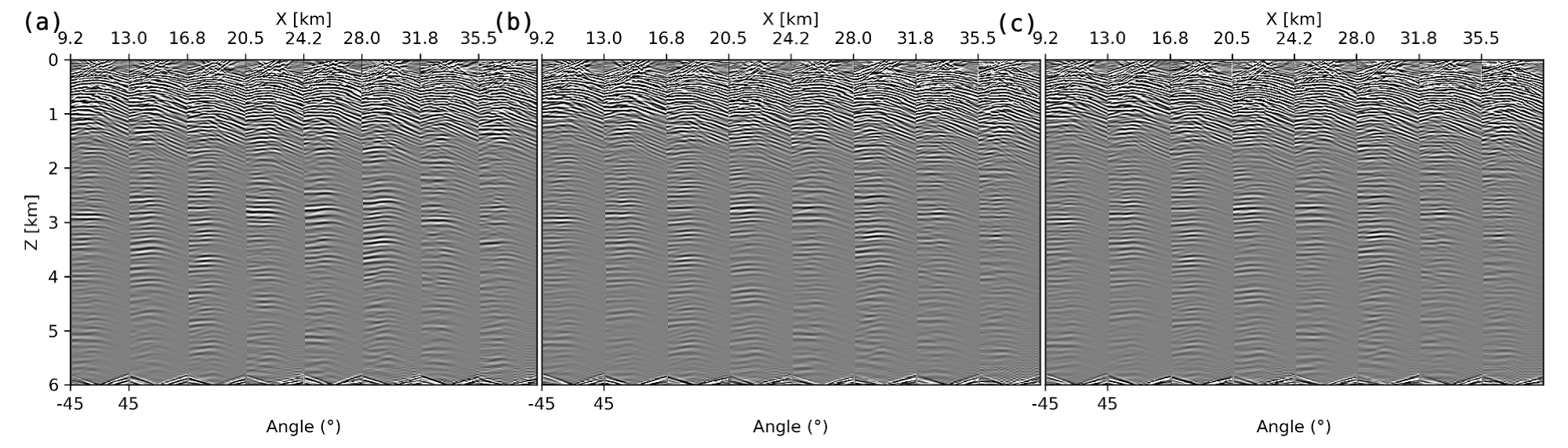} 
\caption{Angle-domain common image gathers (ADCIGs) computed using (a) the initial velocity model, (b) the SoftDTW inverted model, and (c) the inverted model from our misfit.}
\label{fig:chevron_adcig} 
\end{figure}

Figure~\ref{fig:chevron_grad} shows the first-iteration FWI gradient for both misfit functions. The gradient are similar in shape and polarity, especially in shallow regions where refractions dominate. The final inverted models superimposed on their RTM images are shown in Figure~\ref{fig:chevron_vp}. Both methods yield comparable results, though our approach produces slightly higher velocities on the left side at 2 km depth. As a final validation, angle domain common image gathers (ADCIGs) for $[-45^\circ,\ 45^\circ]$ were computed using the initial, SoftDTW, and our Time-Shift-NN velocity models (Figure~\ref{fig:chevron_adcig}). The ADCIGs from the inverted models exhibit flatter events than those from the initial model and are nearly identical.
\section{Discussion}
This work introduces a deep learning-based approach to compute trace-wise time shifts required to match two seismic gathers. Since neural networks are numerically differentiable, our approach can be seamlessly integrated within an FWI workflow and has the potential to overcome or at least alleviate the problem of local minima in the high-dimensional landscape of the search space.

A key aspect of our method is training the neural network on realistic time shifts to produce optimal time shifts between the modeled and observed data. Our results demonstrate that the network produces effective time shifts, enabling FWI to converge at a rate comparable to conventional approaches while successfully avoiding cycle skipping, where the $L_2$ norm fails. Our inverted velocity model is of comparable quality to that of the robust SoftDTW algorithm. Moreover, the well log comparison in Figure~\ref{fig:chevron_vp_init}(b) reveals that both methods effectively recover the correct velocities, with our approach showing marginally better accuracy in the shallower region. The primary motivation of our approach is to provide a robust mechanism to initialize FWI without becoming trapped in local minima, thereby delivering a reliable starting model. Once the cycle-skipping barrier is overcome, we can switch to an alternative misfit function, such as the $L_2$ norm, to refine the inversion toward high-resolution models.

The proposed misfit function yields fully differentiable, smoothly behaved gradients, which simplifies integration into the FWI adjoint‐state framework, and it dramatically reduces runtime compared to SoftDTW. On an NVIDIA A100 GPU, our neural network computes adjoint sources in 4.59 s (amortized over 256 shots including training), compared to 521.93 s for our implementation of SoftDTW based on tslearn~\citep{JMLR:v21:20-091}; this corresponds to an 113.7× speed‐up in the adjoint source compute time. On an Intel Cascade Lake–based CPU, Devito’s~\citep{devito-compiler,devito-api} threaded PDE solver runs in 4.31 s per solve (8.62 s per iteration). As a result, SoftDTW’s gradient assembly accounts for 98.38 \% of iteration time, whereas in our method only 34.7 \% of the total iteration time is spent in adjoint‐source computation. Despite these computational gains, inversion performance matches that of SoftDTW, and by framing the misfit as time shift estimation the approach remains transparent and easily adaptable to other waveform‐matching applications. A full timing breakdown is given in Table \ref{tab:misfit_timing_updated}. Our time-shift–NN approach is both faster and more memory-efficient. Specifically, the neural network computes the full adjoint source (321 traces) in a single pass, while SoftDTW is limited to processing only 30 traces at a time on the GPU. This implies a ~10.7× lower per-trace memory footprint for the neural approach.

\begin{table}[ht]
  \centering
  \begin{threeparttable}
    \caption{Adjoint-source vs.\ PDE-solve share by misfit. PDE solves (forward+adjoint) take 8.62~s per iteration (constant).}
    \label{tab:misfit_timing_updated}
    \begin{tabular}{@{}lcccc@{}}
      \toprule
      Misfit                 & Adjoint-source (s) & Total (s) & \% Adjoint-source & \% PDE solves \\
      \midrule
      SoftDTW tslearn & $521.93$     & 530.55     & 89.38\%           & 1.62\%       \\
      Time-Shift-NN                      & 4.59               & 13.21     & 34.72\%           & 65.28\%       \\
      \bottomrule
    \end{tabular}
  \end{threeparttable}
\end{table}
\section{Conclusions}
In this work, we introduced a neural network-based time-shift misfit function for comparing modeled and observed data in FWI and validated its effectiveness using two synthetic datasets. In both cases, the proposed approach achieves accuracy comparable to SoftDTW, particularly in the shallow part of the models, while delivering substantial computational and memory advantages. More broadly, this work underscores the potential of using neural network-based misfit functions as efficient and reliable alternatives to traditional objective functions for large-scale nonlinear inverse problems.
\section*{Acknowledgments}
The authors thank KAUST and DeepWave sponsors for their support.  For computer time, this research leveraged the resources of the Supercomputing Laboratory at KAUST in Thuwal, Saudi Arabia.
\section*{Code Availability}
All data sets used in this study are publicly available. The Marmousi II model can be obtained from the University of Houston’s AGL downloads page (\url{http://www.agl.uh.edu/downloads/downloads.htm}). The Chevron 2014 FWI benchmark data set is available through the SEG workshop open data archive (\url{https://s3.amazonaws.com/open.source.geoscience/open_data/seg_workshop_fwi_2014/seg_workshop_fwi_2014.html}). To reproduce the results, the code is publicly available on GitHub at \url{https://github.com/DeepWave-KAUST/DeepTimeshift-pub}.
\bibliographystyle{unsrtnat}
\bibliography{References}

@article{bunks1995multiscale,
  title={Multiscale seismic waveform inversion},
  author={Bunks, Carey and Saleck, Fatimetou M and Zaleski, S and Chavent, G},
  journal={Geophysics},
  volume={60},
  number={5},
  pages={1457--1473},
  year={1995},
  publisher={Society of Exploration Geophysicists}
}

@article{van2008velocity,
  title={Velocity analysis based on data correlation},
  author={Van Leeuwen, T and Mulder, WA},
  journal={Geophysical Prospecting},
  volume={56},
  number={6},
  pages={791--803},
  year={2008},
  publisher={European Association of Geoscientists \& Engineers}
}

@article{van2010correlation,
  title={A correlation-based misfit criterion for wave-equation traveltime tomography},
  author={Van Leeuwen, T. and Mulder, W. A.},
  journal={Geophysical Journal International},
  volume={182},
  number={3},
  pages={1383--1394},
  year={2010},
  publisher={Blackwell Publishing Ltd Oxford, UK}
}

@inproceedings{luo2011deconvolution,
  title={A deconvolution-based objective function for wave-equation inversion},
  author={Luo, Simon and Sava, Paul},
  booktitle={{SEG International Exposition and Annual Meeting}},
  pages={SEG--2011},
  year={2011},
  organization={SEG}
}

@article{warner2016adaptive,
  title={Adaptive waveform inversion: Theory},
  author={Warner, Michael and Guasch, Llu{\'\i}s},
  journal={Geophysics},
  volume={81},
  number={6},
  pages={R429--R445},
  year={2016},
  publisher={Society of Exploration Geophysicists}
}

@article{zhu2016building,
  title={Building good starting models for full-waveform inversion using adaptive matching filtering misfit},
  author={Zhu, Hejun and Fomel, Sergey},
  journal={Geophysics},
  volume={81},
  number={5},
  pages={U61--U72},
  year={2016},
  publisher={Society of Exploration Geophysicists}
}

@incollection{debens2017full,
  title={Full-bandwidth adaptive waveform inversion at the reservoir},
  author={Debens, Henry A and Mancini, Fabio and Warner, Mike and Guasch, Llu{\'\i}s},
  booktitle={{SEG Technical Program Expanded Abstracts 2017}},
  pages={1378--1382},
  year={2017},
  publisher={Society of Exploration Geophysicists}
}

@article{huang2017full,
  title={Full-waveform inversion via source-receiver extension},
  author={Huang, Guanghui and Nammour, Rami and Symes, William},
  journal={Geophysics},
  volume={82},
  number={3},
  pages={R153--R171},
  year={2017},
  publisher={Society of Exploration Geophysicists}
}

@article{10.1111/j.1365-246X.1991.tb06713.x,
    author = {Luo, Yi and Schuster, Gerard T.},
    title = "{Wave equation inversion of skeletalized geophysical data}",
    journal = {Geophysical Journal International},
    volume = {105},
    number = {2},
    pages = {289-294},
    year = {1991},
    month = {05},
    issn = {0956-540X},
    doi = {10.1111/j.1365-246X.1991.tb06713.x},
    url = {https://doi.org/10.1111/j.1365-246X.1991.tb06713.x},
    eprint = {https://academic.oup.com/gji/article-pdf/105/2/289/2075923/105-2-289.pdf},
}

@article{luo1991wave,
  title={Wave-equation traveltime inversion},
  author={Luo, Yi and Schuster, Gerard T},
  journal={Geophysics},
  volume={56},
  number={5},
  pages={645--653},
  year={1991},
  publisher={Society of Exploration Geophysicists}
}

@article{Bozdag2011845,
	author = {Bozdaǧ, Ebru and Trampert, Jeannot and Tromp, Jeroen},
	title = {Misfit functions for full waveform inversion based on instantaneous phase and envelope measurements},
	year = {2011},
	journal = {Geophysical Journal International},
	volume = {185},
	number = {2},
	pages = {845 – 870},
	doi = {10.1111/j.1365-246X.2011.04970.x},
}

@article{CHI201436,
title = {Full waveform inversion method using envelope objective function without low frequency data},
journal = {Journal of Applied Geophysics},
volume = {109},
pages = {36-46},
year = {2014},
issn = {0926-9851},
doi = {https://doi.org/10.1016/j.jappgeo.2014.07.010},
url = {https://www.sciencedirect.com/science/article/pii/S0926985114002031},
author = {Benxin Chi and Liangguo Dong and Yuzhu Liu},
}

@article{engquist2016optimal,
  title     = {Optimal transport for seismic full waveform inversion},
  author    = {Bj{\"o}rn Engquist and Brittany D. Froese and Yunan Yang},
  journal   = {Communications in Mathematical Sciences},
  volume    = {14},
  number    = {8},
  pages     = {2309--2330},
  year      = {2016},
  publisher = {International Press of Boston, Inc.},
  doi       = {10.4310/CMS.2016.v14.n8.a9},
  url       = {https://doi.org/10.4310/CMS.2016.v14.n8.a9},
}

@article{metivier2016measuring,
  title={Measuring the misfit between seismograms using an optimal transport distance: Application to full waveform inversion},
  author={M{\'e}tivier, Ludovic and Brossier, Romain and M{\'e}rigot, Quentin and Oudet, Edouard and Virieux, Jean},
  journal={Geophysical Supplements to the Monthly Notices of the Royal Astronomical Society},
  volume={205},
  number={1},
  pages={345--377},
  year={2016},
  publisher={The Royal Astronomical Society}
}

@inproceedings{chen2018constructing,
  title={{Constructing misfit function for full waveform inversion based on sliced Wasserstein distance}},
  author={Chen, F and Peter, D},
  booktitle={{80th EAGE Conference and Exhibition 2018}},
  pages={1--5},
  year={2018},
  organization={European Association of Geoscientists \& Engineers}
}

@inproceedings{chen2018misfit,
  title={A misfit function based on entropy regularized optimal transport for full-waveform inversion},
  author={Chen, F and Peter, D},
  booktitle={{SEG Technical Program Expanded Abstracts 2018}},
  pages={1314--1318},
  year={2018},
  organization={Society of Exploration Geophysicists}
}

@article{yang2018application,
  title={{Application of optimal transport and the quadratic Wasserstein metric to full-waveform inversion}},
  author={Yang, Yunan and Engquist, Bj{\"o}rn and Sun, Junzhe and Hamfeldt, Brittany F},
  journal={Geophysics},
  volume={83},
  number={1},
  pages={R43--R62},
  year={2018},
  publisher={Society of Exploration Geophysicists}
}

@article{ramos2019long,
  title={{Long-wavelength FWI updates in the presence of cycle skipping}},
  author={Ramos-Mart{\'\i}nez, Jaime and Qiu, Lingyun and Valenciano, Alejandro A and Jiang, Xiaoyan and Chemingui, Nizar},
  journal={The Leading Edge},
  volume={38},
  number={3},
  pages={193--196},
  year={2019},
  publisher={Society of Exploration Geophysicists Tulsa, Oklahoma}
}

@article{kalita2019flux,
  title={Flux-corrected transport for full-waveform inversion},
  author={Kalita, Mahesh and Alkhalifah, Tariq},
  journal={Geophysical Journal International},
  volume={217},
  number={3},
  pages={2147--2164},
  year={2019},
  publisher={Oxford University Press}
}

@article{sakoe1978dynamic,
  title={Dynamic programming algorithm optimization for spoken word recognition},
  author={Sakoe, Hiroaki and Chiba, Seibi},
  journal={IEEE transactions on acoustics, speech, and signal processing},
  volume={26},
  number={1},
  pages={43--49},
  year={1978},
  publisher={IEEE}
}

@article{ma2013wave,
  title={Wave-equation reflection traveltime inversion with dynamic warping and full-waveform inversion},
  author={Ma, Yong and Hale, Dave},
  journal={Geophysics},
  volume={78},
  number={6},
  pages={R223--R233},
  year={2013},
  publisher={Society of Exploration Geophysicists}
}

@article{yang2014using,
  title={Using image warping for time-lapse image domain wavefield tomography},
  author={Yang, Di and Malcolm, Alison and Fehler, Michael},
  journal={Geophysics},
  volume={79},
  number={3},
  pages={WA141--WA151},
  year={2014},
  publisher={Society of Exploration Geophysicists}
}

@inproceedings{chen2021misfit,
  title={Misfit functions based on differentiable dynamic time warping for waveform inversion},
  author={Chen, Fuqiang and Peter, Daniel and Ravasi, Matteo},
  booktitle={{SEG International Exposition and Annual Meeting}},
  pages={D011S029R002},
  year={2021},
  organization={SEG}
}

@inproceedings{cuturi2017soft,
  title={Soft-dtw: a differentiable loss function for time-series},
  author={Cuturi, Marco and Blondel, Mathieu},
  booktitle={International conference on machine learning},
  pages={894--903},
  year={2017},
  organization={PMLR}
}

@article{chen2022cycle,
  title={Cycle-skipping mitigation using misfit measurements based on differentiable dynamic time warping},
  author={Chen, Fuqiang and Peter, Daniel and Ravasi, Matteo},
  journal={Geophysics},
  volume={87},
  number={4},
  pages={R325--R335},
  year={2022},
  publisher={Society of Exploration Geophysicists}
}

@inproceedings{kalita2023soft,
  title={Soft-dynamic time warping divergence as a misfit measure in full-waveform inversion},
  author={Kalita, M and Purcell, C and Casasanta, L},
  booktitle={{84th EAGE Annual Conference \& Exhibition}},
  pages={1--5},
  year={2023},
  organization={European Association of Geoscientists \& Engineers}
}

@inproceedings{blondel2021differentiable,
  title={Differentiable divergences between time series},
  author={Blondel, Mathieu and Mensch, Arthur and Vert, Jean-Philippe},
  booktitle={International Conference on Artificial Intelligence and Statistics},
  pages={3853--3861},
  year={2021},
  organization={PMLR}
}

@article{9444561,
  author={Dramsch, Jesper Sören and Christensen, Anders Nymark and MacBeth, Colin and Lüthje, Mikael},
  journal={IEEE Transactions on Geoscience and Remote Sensing}, 
  title={Deep Unsupervised 4-D Seismic 3-D Time-Shift Estimation With Convolutional Neural Networks}, 
  year={2022},
  volume={60},
  number={},
  pages={1-16},
  doi={10.1109/TGRS.2021.3081516}
}

@article{10.1190/geo2019-0724.1,
    author = {Dhara, Arnab and Bagaini, Claudio},
    title = "{Seismic image registration using multiscale convolutional neural networks}",
    journal = {Geophysics},
    volume = {85},
    number = {6},
    pages = {V425-V441},
    year = {2020},
    month = {10},
    issn = {0016-8033},
    doi = {10.1190/geo2019-0724.1},
    url = {https://doi.org/10.1190/geo2019-0724.1},
    eprint = {https://pubs.geoscienceworld.org/geophysics/article-pdf/85/6/V425/5220796/geo-2019-0724.1.pdf},
}

@inbook{doi:10.1190/segam2020-3427292.1,
author = {Zhun Li and Aria Abubakar},
title = {Complete sequence stratigraphy from seismic optical flow without human labeling},
booktitle = {{SEG Technical Program Expanded Abstracts 2020}},
chapter = {},
pages = {1248--1252},
year = {2020},
doi = {10.1190/segam2020-3427292.1},
URL = {https://library.seg.org/doi/abs/10.1190/segam2020-3427292.1},
eprint = {https://library.seg.org/doi/pdf/10.1190/segam2020-3427292.1},
publisher = {Society of Exploration Geophysicists}
}

@inproceedings{ronneberger2015u,
  title={U-net: Convolutional networks for biomedical image segmentation},
  author={Ronneberger, Olaf and Fischer, Philipp and Brox, Thomas},
  booktitle={{Medical Image Computing and Computer-Assisted Intervention--MICCAI 2015: 18th International Conference, Munich, Germany, October 5-9, 2015, Proceedings, Part III 18}},
  pages={234--241},
  year={2015},
  organization={Springer}
}

@article{plessix2006review,
  title={A review of the adjoint-state method for computing the gradient of a functional with geophysical applications},
  author={Plessix, R-E},
  journal={Geophysical Journal International},
  volume={167},
  number={2},
  pages={495--503},
  year={2006},
  publisher={Blackwell Publishing Ltd Oxford, UK}
}

@article{liu1989limited,
  title={{On the limited memory BFGS method for large scale optimization}},
  author={Liu, Dong C and Nocedal, Jorge},
  journal={Mathematical programming},
  volume={45},
  number={1-3},
  pages={503--528},
  year={1989},
  publisher={Springer}
}

@article{devito-compiler,
  author = {Luporini, Fabio and Louboutin, Mathias and Lange, Michael and Kukreja, Navjot and Witte, Philipp and H\"{u}ckelheim, Jan and Yount, Charles and Kelly, Paul H. J. and Herrmann, Felix J. and Gorman, Gerard J.},
  title = {{Architecture and Performance of Devito, a System for Automated Stencil Computation}},
  year = {2020},
  issue_date = {March 2020},
  publisher = {Association for Computing Machinery},
  address = {New York, NY, USA},
  volume = {46},
  number = {1},
  issn = {0098-3500},
  url = {https://doi.org/10.1145/3374916},
  doi = {10.1145/3374916},
  journal = {ACM Trans. Math. Softw.},
  month = {apr},
  articleno = {6},
  numpages = {28}
}

@article{devito-api,
  author = {Louboutin, M. and Lange, M. and Luporini, F. and Kukreja, N. and Witte, P. A. and Herrmann, F. J. and Velesko, P. and Gorman, G. J.},
  title = {Devito (v3.1.0): an embedded domain-specific language for finite differences and geophysical exploration},
  journal = {Geoscientific Model Development},
  volume = {12},
  year = {2019},
  number = {3},
  pages = {1165--1187},
  url = {https://www.geosci-model-dev.net/12/1165/2019/},
  doi = {10.5194/gmd-12-1165-2019}
}

@article{JMLR:v21:20-091,
  author  = {Romain Tavenard and Johann Faouzi and Gilles Vandewiele and
             Felix Divo and Guillaume Androz and Chester Holtz and
             Marie Payne and Roman Yurchak and Marc Ru{\ss}wurm and
             Kushal Kolar and Eli Woods},
  title   = {Tslearn, A Machine Learning Toolkit for Time Series Data},
  journal = {Journal of Machine Learning Research},
  year    = {2020},
  volume  = {21},
  number  = {118},
  pages   = {1-6},
  url     = {http://jmlr.org/papers/v21/20-091.html}
}

@article{doi:10.1190/1.2172306,
author = {Gary S. Martin and Robert Wiley and Kurt J. Marfurt},
title = {Marmousi2: An elastic upgrade for Marmousi},
journal = {The Leading Edge},
volume = {25},
number = {2},
pages = {156-166},
year = {2006},
doi = {10.1190/1.2172306},
URL = {https://doi.org/10.1190/1.2172306},
eprint = { https://doi.org/10.1190/1.2172306}
}

\end{document}